\documentclass[fleqn,usenatbib]{mnras}

\usepackage{newtxtext,newtxmath}

\usepackage[T1]{fontenc}

\DeclareRobustCommand{\VAN}[3]{#2}
\let\VANthebibliography\thebibliography
\def\thebibliography{\DeclareRobustCommand{\VAN}[3]{##3}\VANthebibliography}

\usepackage{graphicx}	
\usepackage{amsmath}	
\usepackage{siunitx}
\usepackage{subcaption}
\usepackage{caption}

\newcommand{\Mi}{ $M_{\text{ZAMS}}$ } 
 
\newcommand{\aOv}{ $\alpha_{\text{ov}}$ } 
\newcommand{\OmOmC}{ $\Omega / \Omega_{\text{crit}}$ } 
\newcommand{\aSC}{ $\alpha_{\text{sc}}$ } 

\title[Rotation and the BH PI boundary at low Z]{The black hole - pair instability boundary for high stellar rotation}

\author[E. R. J. Winch et al.]{
Ethan R. J. Winch,$^{1,2}$\thanks{E-mail: ethan.winch@armagh.ac.uk}
Gautham N. Sabhahit,$^{2}$
Jorick S. Vink,$^2$
and Erin R. Higgins$^{2,1}$
\\
$^{1}$School of Maths and Physics, Queen's University Belfast, Northern Ireland, University Road, BT7 1NN\\
$^{2}$Armagh Observatory and Planetarium (AOP),
              Armagh, College Hill, BT61 9DB
}

\date{Accepted 2025 April 23. Received 2025 April 23; in original form 2025 March 14}

\pubyear{2025}

\begin{document}
\label{firstpage}
\pagerange{\pageref{firstpage}--\pageref{lastpage}}
\maketitle

\begin{abstract}
The Pair Instability (PI) boundary is crucial for understanding heavy merging Black Holes (BHs) and the second mass gap's role in galactic chemical evolution. So far, no works have critically and systematically examined how rotation and mass loss affect the PI boundary or BH masses below it. Rapid rotation significantly alters stellar structure and mass loss, which is expected to have significant effects on the evolution of stellar models. We have previously derived a critical core mass independent of stellar evolution parameters, finding the BH (Pulsational) PI boundary at $M_{\rm CO, crit} = 36.3 M_\odot$ for a carbon-oxygen (CO) core. Using MESA, we model massive stars around the PI boundary for varying rotation rates and metallicities. We implement mechanical mass loss in MESA, studying its effects on massive stars in low-metallicity environments. Below $1/100$th $Z_\odot$, mechanical mass loss dominates over radiative winds. We check the BH-PI boundary for rapid rotators to confirm our critical core mass criterion and derive model fits describing rotation’s impact on core and final masses. Fast rotators reach a point (typically $\Omega / \Omega_{\rm crit} \approx 0.6$) where the entire star becomes chemically homogeneous, evolving like a stripped star. This lowers the maximum BH mass before PI to its critical core mass of $M_{\rm CO, crit} = 36.3 M_\odot$, aligning with the bump feature in the BH mass distribution observed by LIGO/VIRGO.
\end{abstract}

\begin{keywords}
stars: massive -- stars: black holes -- stars: evolution -- stars: Population II 
\end{keywords}



\section{Introduction}

The end-point of a massive star may not only involve a neutron star or black hole (BH), but could also lead to a disruptive pair instability (PI) supernova (PISN), leaving no stellar remnant at all \citep[e.g.][]{Heger03,Woosley17}. The first mass gap between neutron stars and BHs may be found in the range 2-5 $M_\odot$
\citep[][]{Bailyn98, Ozel10}, but there is also thought to be a second mass gap caused by PI \citep{Woosley17,Farmer19,Renzo20_LowerPISN}.

Until the BH merger event of GW 190521 involving an 85 and 66 $M_\odot$ BH as detected by LIGO-Virgo \citep[][]{Abbott20}, this second gap was thought to be in the BH mass range of roughly $50-120 M_\odot$ \citep[][]{Woosley17,Marchant19,Farmer19}, but this was based on the assumption that all of the most massive stars remove their entire hydrogen (H) envelope. However, the fact that massive stars might be able to maintain (part of) this H envelope at low metallicity ($Z$) due to weaker wind mass loss had been overlooked. 
\textcolor{black}{\citet{Liu2020, vink21, Farrell21, Kinugawa21} placed the observation of GW190521 in the context of low metallicity ($Z$) stars.} 
In addition, the interior assumptions on the amount of mixing, characterised by the core overshooting, had not been fully considered \textcolor{black}{in a stellar evolution context \citep{vink21,Tanikawa21Overshoot}. }
\textcolor{black}Furthermore, there is some uncertainty in the $C12(\alpha , \gamma)O16$ reaction rate \citep{Takahashi18,Farmer20,Costa21}.

In two previous papers \citep[][]{vink21, Winch24} we showed that stars in the initial mass range $90-100 M_\odot$ are capable of producing core masses below the critical limit of $M_{\rm CO, crit} = 36.3 M_\odot$ thus avoiding PI while also retaining the H-rich envelope due to weaker winds at low Z. Thus potentially producing black holes up to $93.3 M_\odot$. \citet{Winch24} specifically investigated a set of stellar evolution assumptions which were most likely to affect the evolution of massive stars and the location of the PI boundary using the global stability criterion of \citet{Stothers99}. 
The boundary between BH and PI was found to be characterised by a constant critical core mass of $M_{\rm CO, crit} = 36.3 M_\odot$ (for the various physical mechanisms tested therein). That is, models with a core mass above this constant critical core mass underwent PI while models below it directly formed black holes. This core mass was determined to be separate from the total mass, as it is the PI occurring in the core which leads to a collapse, thus it is possible to expect larger black hole progenitors with H-rich stars as \citet{Fernandez18} shows that blue supergiants (BSGs) lose on the order of $0.1-1 M_\odot$ (their Figure 6). We then computed a grid of models which populated our parameter space of stellar evolution assumptions, allowing us to define relationships for the parameters which had the most significant effect on the final mass and core mass, and produce a population examining the most massive possible black holes before PI sets in.

In \cite{Winch24}, we probed the robustness of the pair-instability boundary using low metallicity very massive star (VMS) models with rotation rates up to 40\% of the critical rotation. One major issue we encountered in that work was the potential for models to spin up to near- or super-critical rotation, especially at low $Z$, where the models lose very little angular momentum due to weaker radiation-driven wind mass loss. A rapid spin-up typically occurs at the end of core-H burning where the model undergoes a total contraction phase. In this work, we focus on the regime of rapidly rotating, low $Z$ stars, and their impact on the pair-instability boundary.

Rotation undeniably plays a significant role in massive star evolution, and while the principal mechanisms of rotation in stellar evolution are understood, there are significant differences in implementations of rotation effects depending on the particular evolution code used, initial assumptions, or stellar physics implementations.

In addition to rotation, there is also the question of the Spruit-Tayler (ST) Dynamo and whether it should be included in stellar models. Around $\sim 7 \%$ of massive stars have observed magnetic fields \citep{wade16}, though these are external magnetic fields and a similar statistic is unknown for the internal magnetic fields. The Bonn Evolution Code \citep[e.g.][]{Yoon12} makes use of the ST Dynamo while GENEC does not \citep[e.g. ][]{Ekstrom08}. There can be significant differences between models calculated at Population III metallicity in the aforementioned works - for example, in the final mass. The ongoing question about these differences concerns the origin. Given the nature of the models, the main differences are in how rotation is handled between MESA/Bonn (which use a similar implementation) and Geneva, the inclusion of a Spruit-Tayler Dynamo, or the method of calculating mass loss for rapidly rotating low metallicity massive stars. Our brief investigation into this is discussed in our results section.

Significant work has been performed regarding the effects of rotation on stellar evolution \citep{Meynet97, Maeder00B, Meynet06, Georgy13, Georgy21}, highlighting the importance of rotation - especially at low metallicity \citep{Sibony24}, and for stars which are expected to experience mechanical mass loss events such as in \citet{Meynet06}. In mechanical mass loss, the star reaches the critical rotation velocity. At this point, the balance of forces at the surface ($\textbf{g}_{\rm rad} + \textbf{g}_{\rm rot} + \textbf{g}_{\rm grav} = 0$) becomes unstable. That is to say that $\textbf{g}_{\rm rad} + \textbf{g}_{\rm rot} + \textbf{g}_{\rm grav} > 0$, where the star's surface layers become unbound and the star as a whole can be said to break up. This then leads to large amounts of mass loss, typically for a short period before it settles back to a sub-critical state, though it may become critical again later on. 

While we previously considered the effects of overshooting, semi-convection, and 
rotation, the parameter space of rapid rotation has not been explored. \citet{Marchant20} studied the effect of rapid rotation on the PI criterion directly, but the full effects of stellar evolution in terms of the interplay of rotation and mass loss have yet to be studied.

There are still many uncertainties in the interplay between rotation and mass loss ($\dot{M}$), and it is in fact not even established whether rotation should always lead to an {\it increase} in $\dot{M}$ but in certain parameter spaces could result in {\it decreases} \citep[][]{Muller14}. 
In stellar codes such as MESA the default implementation is that of \citet{Friend86, Langer97, Yoon10} where the mass-loss rate increases rapidly due to the approach of break-up. In this implementation, the radiative mass-loss rate is directly adjusted by a factor of rotation. As the rotation rate gets closer to the critical one, this enhancement would theoretically push up the rotation rate continuously to stop the model becoming supercritical. For very massive stars the effect of the approach of the Eddington limit is probably the more dominant factor \citep[][]{Glatzel98}, and these effects were unified by \citet{Maeder00B}, and expressed in terms of the the $\Omega \Gamma$ limit. 

Mass loss boosts including this $\Omega \Gamma$ limit have mostly been confined to models computed with the GENEC code, but were also included in MESA in the mass-loss framework for low metallicity (Z) by \citet{Sab23} and are refined in the current paper.
For massive, and very massive stars, rotation can change the final fate of the star in ways which are non-linear - a star that has a rapid rotation rate may lose the majority of its angular momentum by the end of its evolution, alternatively a moderate rotator may spin up by the end of its evolution \citet{Meynet02}. 
While PISN and Pulsational Pair Instability Supernovae (PPISN) do not rely on rotation for formation channels, there are a few studies which take into account how rotation would change the distribution of PI occurrences such as \citet{Gabrielli24}.

In this work we complete our examination of the parameter space by focusing on rapid rotation at low metallicity, completing the investigation for H-rich massive stars started in \citet{vink21}. We achieve this in the following manner; in Section \ref{s-Methods}, we provide details on our MESA implementation, and focus on how rotation is treated within MESA, before giving details as to how we resolve supercritical rotation rates. In Section \ref{s-Results}, we apply these methods to a series of H-rich massive star models and explore the effect of rotation at various masses and metallicities, and in the presence or lack of a Spruit-Tayler dynamo. Section \ref{s-Discussion} allows us to examine our work in the context of others, including other methods of handling supercritical rotation and methods of mass loss eruptions, while we finish with our conclusions in Section \ref{s-Conclusions}. Finally, our Appendix \ref{Ap-TableOfModels} provides a list of the models we use.

Table \ref{T-RotationDefs} outlines our key physical mechanisms described throughout this paper for clarity when comparing to other works. 
\begin{table}
\centering
\begin{tabular}{c | l}
\hline
Quantity & Definition \\
\hline
\hline
$R$ & Radius of the star assuming a spherical volume \\
$\varv_{\rm rot}$ & Rotational velocity in km/s \\
$\Omega$ & Angular rotational velocity in rad/s \\
$\varv_{\rm crit}$ & Critical rotation in MESA in km/s \\
$\Omega_{\rm crit}$ & Critical rotation used in MESA in rad/s \\
\OmOmC & Fraction of critical velocity \\
\hline
$\varv_{\rm crit, 1, MM}$ & First critical velocity in km/s \citep[][]{Maeder00B}\\
$\varv_{\rm crit, 2, MM}$ & Second critical velocity in km/s \citep[][]{Maeder00B}\\
$\Omega_{\rm crit, MM}$ & Critical rotation from \citet{Maeder00B} in rad/s \\
$\omega$ & Fraction of angular velocity at \\ & \hspace{3pt} break up \citep[from][]{Maeder00B} \\
$R_{\rm e}$ & Equatorial radius of the star \citep[from][]{Maeder00B} \\
$R_{\rm p}$ & Polar radius of the star \citep[from][]{Maeder00B} \\
$R_{\rm eb}$ & $R_{\rm eq}$ at break-up \\
$R_{\rm pb}$ & $R_{\rm p}$ at break up \\
$\Gamma_{\Omega}$ & Eddington limit for a rotating star \\ 
\hline
\end{tabular}
\caption{Definition of all terms used in this work for rotation. All quantities below the line are directly related to \citet{Maeder00B}, while quantities above the line are for MESA. If unspecified in the text, the quantity belongs to the MESA definition.}
\label{T-RotationDefs}
\end{table}

\section{Methods}
\label{s-Methods}

We first give a brief overview of the stellar evolution code used and the input parameters in Section \ref{ss-Methods_InitParams}, and our mixing implementation in Section \ref{ss-Methods_Mixing}. We then provide a detailed description of our mass loss implementation to deal with rotation-induced mass loss and possible mechanical mass loss during super-critical rotation phases (Section \ref{ss-Methods_Mdot}). Finally we discuss PI and our $M_{\rm crit}$ experiment in Section \ref{ss-Methods_PI_MCrit}.

\subsection{Initial mass and metallicity}
\label{ss-Methods_InitParams}

We use the MESA stellar evolution code \citep[version r15140;][]{Paxton11,Paxton13,Paxton15,Paxton18,Paxton19} to calculate our model grid. Our initial mass ranges from $50 - 100 M_{\odot}$. Initial $Z$ describes the abundance of elements other than H and He. The initial $Z$ in our models are $Z = 0.002, 0.0002$ to $0.00002$. The $Z$ range should roughly span a tenth to a thousandth solar metallicity, and goes down to near-primordial metallicities. The individual metal mass fraction spread follows solar-scaled ratios from \citet{Grevesse98}. The H and He mass fractions are then obtained as follows. Firstly the He mass fraction Y is determined as $Y = Y_{\rm prim} + (\Delta Y / \Delta Z) \times Z$, where the primordial He mass fraction is $Y_{\rm prim} = 0.24$ and the enrichment factor is $(\Delta Y / \Delta Z) = 2$. The H mass fraction is then obtained by satisfying the unity condition: $X + Y + Z = 1$.

\subsection{Mixing processes: convection, overshooting and rotation}
\label{ss-Methods_Mixing}

Convection is treated in our models using the standard Mixing Length Theory (MLT) from \citet{Cox68} with an efficiency parameter of $\alpha_{\rm MLT} = 1.82$ as in \citet{Choi16}. We also employ the Ledoux criterion which takes into account the chemical gradients when checking for convective stability of a layer. Such chemical gradients can stabilise layers against convection, which can give rise to extended semi-convective regions, for example above the receding convective core during the MS phase of massive stars.

In order to capture convective boundary mixing, we use a step overshooting prescription with the free parameter $\alpha_{\rm ov} = 0.1$, which is consistent with our \cite{Winch24} $M_{\rm crit}$ experiment fiducial model ($90 M_\odot$, $1/10$th $Z_\odot$, $\Omega =0.2 \Omega_{\rm crit}$, $\alpha_{\rm ov} = 0.1$, and \textcolor{black}{the semiconvective efficiency parameter} $\alpha_{\rm sc} = 0.1$) and will also provide for us the largest black holes. Overshooting introduces additional mixing above the convective core and captures mixing introduced by fluid elements that have over-shot the formal convective boundary defined by the standard Schwarzschild convective criterion.  We also employ the following rotational-induced mixing instabilities in our models--Solberg-Hoiland, Secular Shear, Eddington-Sweet circulation, and Goldreich-Schubert-Fricke--using a diffusive approach following \citet{Heger00}. We also test the effects of internal magnetic fields on the core-envelope coupling in the presence of rapid rotation by using the Spruit-Tayler dynamo \citep{Spruit02, Heger05, Petrovic05} implemented in MESA. 

Our models employ the MLT++ option in MESA when dealing with super-adiabatic convective layers, which in the presence of super-Eddington conditions can result in very low density, inflated envelopes. The reasons we use MLT++ are two-fold. Multi-D models have hinted towards such layers becoming porous through which photons can preferentially escape, which could potentially destroy such inflated layers \citep[][]{Goldberg2022}. Second, models encounter numerical problems when dealing with such layers, with each timestep getting infinitesimally smaller. For details about MLT++, see \citet{Paxton13}.

As we are focused on fast rotators, a significant number of our models become chemically homogenous. The physics of rotationally-induced chemically homogenous evolution is still under debate, and requires very efficient rotational mixing. However there is no empirical evidence that low metallicity WR stars rotate any more rapidly than higher metallicity counterparts \citep{Vink17Harries}.

\subsection{Mass loss}
\label{ss-Methods_Mdot}

The intense radiation pressure of massive stars are capable of driving and sustaining a radial outflow which we call a wind. This occurs when momentum transfer occurs from the radiation field to resonance lines of specific ions, which is further augmented by the Doppler effect associated with the outward flow \citep[for a review, see ][]{Puls08, Vink22}.  Enormous amounts of mass can be lost via such line-driven winds, especially at high $Z$ where the momentum transfer can occur across millions of iron lines. At low $Z$, the mass loss scales down due to the lower amount of iron available for wind driving. 

\begin{figure}
    \centering
    \includegraphics[width=1.05\linewidth]{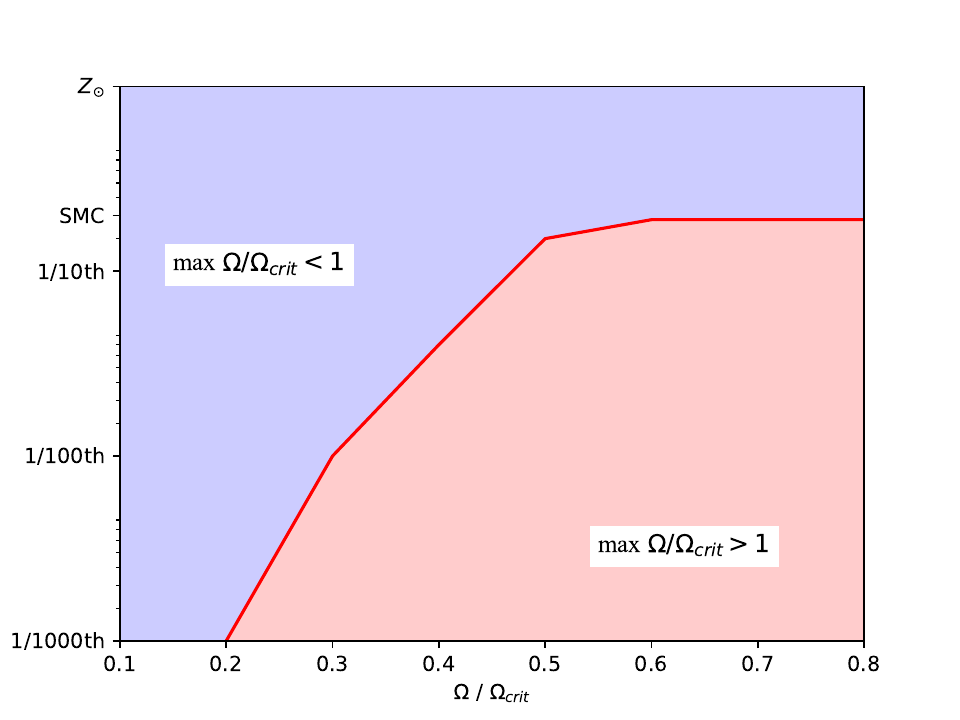}
    \caption{Sketch of where mechanical mass loss is relevant for models. Models in the blue region have a maximum $\Omega / \Omega_{\rm crit} < 1$, meaning that these models do not require mechanical mass loss, while models in the red region have a maximum $\Omega / \Omega_{\rm crit} > 1$. Adapted and extended from \citet{Winch24}.}
    \label{Fig-RotationSketch}
\end{figure}

Typically during the MS, the star gradually increases in radius evolving towards cooler temperatures, and by doing so, the star spins down and it remains below the break-up velocity. In Figure \ref{Fig-RotationSketch}, we show a schematic of the maximum ratio of angular rotational velocity, $\Omega$ in $\mathrm{rad\,s^{-1}}$ to the critical angular velocity, $\Omega_\mathrm{crit}$ as defined in our MESA models in metallicity and initial rotation space. In the case of high $Z$ and low rotation, the star spins down. However, the combination of low $Z$ and rapid rotation can result in a rapid spin-up of a star ultimately reaching break-up. In Nature, when such near-critical velocities are encountered, the star likely ejects the material leading to a mechanical wind. 

The aim of this sub-section is to detail our implementation of mass loss and how we treat near break-up situations in our VMS models. The mass-loss prescription used in our models can be broken down as follows: radiation-driven wind mass loss only, rotationally enhanced radiation-driven mass loss when approaching the so-called $\Omega \Gamma$ limit, and mechanical mass loss for models that cross the $\Omega \Gamma$ limit.

\subsubsection{Radiation-driven wind mass loss, $\dot{M}_\mathrm{rad}(\Omega=0)$}
\label{sss-Methods_RadMdot}

For radiative-driven wind mass loss in the non-rotating case, we use the mass-loss framework developed in \citet{Sab23} which was specifically tailored to study VMSs at low $Z$. The framework consists of the following physical prescriptions. For surface effective temperatures between 4000 K and 100,000 K, we use the mass-loss kink formalism by \citet{Vink11}, comprising of the optically-thin O-star rates from \citet{Vink01} which transitions to an optically-thick mass loss scaling from \citet{Vink11} via a kink. Above the kink, the mass-loss rate scales steeply with the Eddington parameter $\Gamma_\mathrm{Edd}$ with a power law slope of roughly 5, where
\begin{equation}
    \label{Eq-GammaEdd}
    \Gamma_{\rm Edd} = \frac{L}{L_{\rm Edd}} = \frac{\kappa L }{4 \pi c G M}
\end{equation}

For temperatures cooler than 4000 K, we use the red supergiant wind prescription from \citet{deJager88}, and for temperatures greater than 100,000 K, we use the maximum of the absolute rates predicted by \citet{Sander20} and \citet{Vink17} \footnote{The original mass loss framework in \citet{Sab23} only used \citet{Sander20} when the temperatures increased above 100,000 K. However, we realised that at very low $Z$, the rates predicted by \citet{Sander20} becomes unrealistically small. So we have added the mass-loss prescription for stripped stars by \citet{Vink17} as our minimum floor.}.

\subsubsection{Rotation-enhanced radiation-driven mass loss}
\label{sss-Methods_RotBoostMdot}
 
In this work, we enhance the radiation-driven wind mass loss for models close to their so-called $\Omega \Gamma$ limit following \citet{Maeder00B}, who developed a rotation boost for the radiation-driven mass-loss by combining the radiative-driven wind theory \citep[][]{Castor75, Pauldrach86, Kudritzki89, Puls96} and the von Zeipel theorem which takes into account the gravity-darkening effects \citep{vonZeipel24, Tassoul78}. 

\citet{Sab23} tested the rotation-enhanced mass-loss boost by \citet{Maeder00B} in MESA. However, an approximation was made in the formula that is only valid for velocities below approximately 70–80\% of the critical velocity, which was sufficient for the purposes of the paper. The approximation no longer holds for the velocities encountered in this study. Below, we go into details regarding the rotation-enhanced mass-loss boost from \citet{Maeder00B}, and the relaxation of the approximation made in \citet{Sab23}.

\citet{Maeder00B} showed that two separate roots exist for break-up when $\textbf{g}_\textbf{tot} = \textbf{g}_\textbf{rad} + \textbf{g}_\textbf{rot} + \textbf{g}_\textbf{grav} = \textbf{0}$, i.e.,  when the vector sum of the radiative, rotational and Newtonian gravity forces equals zero. The first root, $\varv_\mathrm{crit,1}$, occurs for low Eddington parameters below $\sim 0.64$, when $\textbf{g}_\textbf{eff} = \textbf{g}_\textbf{rot} + \textbf{g}_\textbf{grav} = \textbf{0}$. The first critical velocity is given by
\begin{equation}
    \label{Eq-MM00_vcrit_1}
    \begin{split}
    \varv_{\rm crit, 1, MM} = \sqrt{ \frac{G M }{R_{\rm eb}} } = \sqrt{ \frac{2}{3} \frac{G M }{R_{\rm pb}} }, \\
    {\Omega_\mathrm{crit, MM} = \sqrt{\frac{GM}{R_\mathrm{eb}^3}} = \sqrt{\frac{8}{27}\frac{GM}{R_\mathrm{pb}^3}}    }    
    \end{split}
\end{equation}
where $R_\mathrm{eb}$ and $R_\mathrm{pb}$ are equatorial and polar radii at break-up, which are related by a factor of 1.5. While rapid rotation can oblately distort the structure of the star, with the equatorial radius changing significantly with rotation, the polar radius remains roughly constant with maximum deviation of 5\%. We therefore make an approximation when calculating the first critical velocity, that is, we set the break-up polar radius equal to the non-rotating radius, $R_\mathrm{pb}\approx R_\mathrm{p}(\Omega=0)$.

For Eddington parameters greater than $\sim 0.64$, a different root is reached first. The star can reach break-up for velocities much lower than the first critical velocity. The relevant root is captured using the so-called $\Gamma_\Omega$ parameter, which is the appropriate Eddington parameter which depends on the angular rotation:
\begin{equation}
\label{Eq-GammaOmega}
    \Gamma_{\Omega} = \frac{ \Gamma_{\rm Edd} }{ 1 - \frac{\Omega^2}{2 \pi G \rho_m} }
\end{equation}
Here $\rho_m$ is the average density of the model taking into account the oblate-distortion effect, i.e., $\rho_m = M/V(\Omega)$.  In order to find the critical solution, we need to find the critical $\Omega$ which satisfies the condition $\Gamma_{\Omega} = 1$ for a given $\Gamma_\mathrm{Edd}$. The solution is not trivial because the volume $V$ itself is a function of rotation. 

To proceed, we calculate the volume by solving the equation of the isobaric surface under shellular approximation \citep[see Appendix of][for the equation]{Maeder00B}. We do this numerically by choosing values of $\Omega$, calculating the volume and checking whether $\Gamma_{\Omega} = 1$ is satisfied. We start from zero $\Omega$ and increase upwards until $\Gamma_{\Omega} = 1$ is satisfied. This iteration to search the critical $\Omega$ is performed at every timestep in our MESA model. The second critical velocity is given by
\begin{equation}
    \label{Eq-MM00_vcrit_2}
    \begin{split}
    & \varv_{\rm crit, 2, MM}^2 = \frac{81}{16} \frac{1 - \Gamma_{\rm Edd}}{V'(\omega)} R_{\rm e}^2 (\omega)\\ &
    = \frac{9}{4} \varv_{\rm crit, 1, MM}^2 \frac{1 - \Gamma_{\rm Edd}}{V'(\omega)} \frac{R_{\rm e}^2 (\omega)}{R_{\rm pb}^2}
    \end{split}
\end{equation}
where \textcolor{black}{$R_\mathrm{e}$ is the equatorial radius of the star at a specific time, and} $\omega$ is the angular velocity in terms of the classical break-up angular velocity obtained from Eq. \ref{Eq-MM00_vcrit_1}, given by 
\begin{equation}
\label{Eq-omegaSquared}
    \omega^2 = \frac{\Omega^2}{\Omega_\mathrm{crit,MM}^2} = \dfrac{\Omega^2 }{8GM/27R_\mathrm{pb}^3}.
\end{equation}

\begin{figure}
    \centering
    \includegraphics[width=1.05\linewidth]{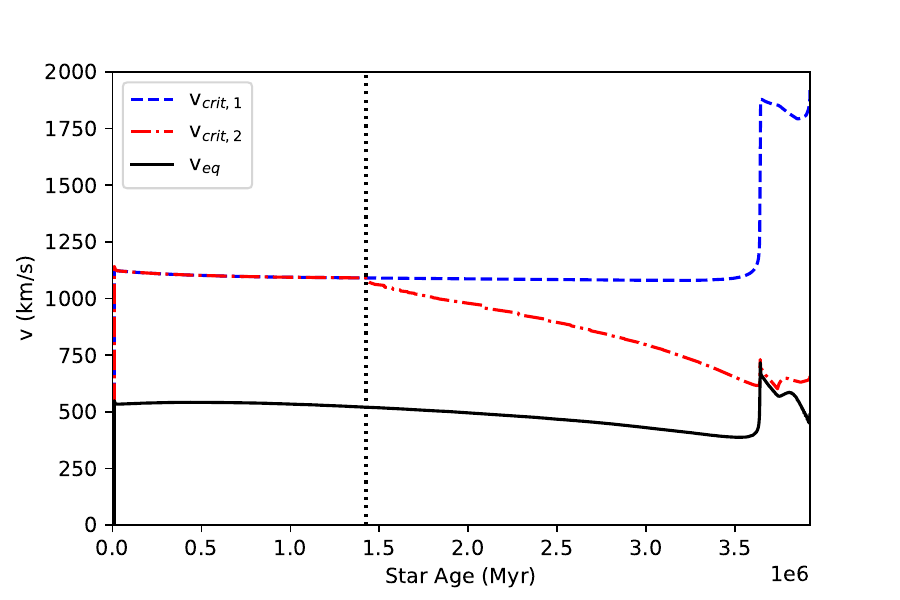}
    \caption{Critical velocity evolution of a $100 M_{\odot}$ model with an initial rotation rate $\Omega / \Omega_{\rm crit} = 0.6$ and a metallicity of $Z = 1/100 Z_\odot$. The critical velocities $\varv_{\rm crit, 1}$ and $\varv_{\rm crit, 2}$ defined in equations \ref{Eq-MM00_vcrit_1} and \ref{Eq-MM00_vcrit_2} are shown with the dashed blue, and dash-dot red lines respectively. The corresponding equatorial velocity of the model, $\varv_{\rm eq}$ is shown by the black solid line. The dotted black vertical line shows the location where $\Gamma_{\rm Edd} > 0.64$, which triggers the separation of $\varv_{\rm crit, 1}$ and $\varv_{\rm crit, 2}$ as the assumption that these values are similar is no longer accurate. The model is seen to exceed $\varv_{\rm crit, 2}$ at $\sim 3.5$ Myr.}
    \label{Fig-100M_06Om_Vcrit_plot}
\end{figure}

In Figure \ref{Fig-100M_06Om_Vcrit_plot}, we show the first and second critical velocities calculated at each timestep along with the model equatorial velocity. The black dashed line marks the location when the Eddington parameter crosses roughly 0.64, and the second critical velocity becomes the more relevant root of the force balance equation. In fact, for Eddington parameter less than 0.64, no root exists for $\Gamma_{\Omega} = 1$ and the only root is the first one. We therefore see a bifurcation in the critical velocities at roughly 1.5 Myr as the model $\Gamma_\mathrm{Edd}$ crosses 0.64.

For radiative-driven wind mass loss in the presence of rotation, we use the following mass loss prescription
\begin{equation}
    \label{Eq-MM00_MdotBoost_Eqn}
    { \dot M_\mathrm{rad} (\Omega) = f_\mathrm{rot,boost,MM}\times{ \dot M_\mathrm{rad} (0)} }
\end{equation}
with the mass loss rotation boost factor from \citet{Maeder00B}.  
\begin{equation}
    \label{Eq-MM00_MdotBoost}
    f_\mathrm{rot,boost, MM} = \frac{ (1 - \Gamma)^{\frac{1}{\alpha} - 1} }{ \left[ 1 - \frac{\Omega^2}{2 \pi \rho_{m} G} - \Gamma \right]^{\frac{1}{\alpha} - 1} }
\end{equation}

Here, $\alpha$ is the CAK force multiplier which is fixed to $\alpha = 0.52$ given the high temperatures of our models \citep{Lamers95}. 

A very similar rotation boost factor was implemented in \citet{Sab23}. However, the following approximation was made regarding the term $ \frac{\Omega^2}{2 \pi \rho_{m} G} \approx \frac{4v^2}{9v^2_\mathrm{crit,1,MM}}$. We have relaxed this approximation in this work and the $\rho_\mathrm{m}$ of the oblate-distorted model is now calculated every model timestep. This implementation of the \citet{Maeder00B} rotation boost is new in MESA. So it warrants a brief comparison to the default rotation boost factor in MESA, which is a fit from \citet{Langer97} with a modification from \citet{Yoon10} to limit the mass loss to the thermal timescale, $\tau_{\rm KH}$, to wind models from \citet{Friend86}.
The critical velocity in MESA is given by 
\begin{equation}
    \label{Eq-OmegaCrit}
    \varv_{\rm crit} = \sqrt{ \left( 1 - \Gamma_{\rm Edd} \right) \frac{GM}{R}}
\end{equation}
In contrast to the previous critical velocities, the MESA critical velocity does not take the von Zeipel effects into account. The rotation boost is
\begin{equation}
\label{Eq-Langer98_MdotBoost}
    f_\mathrm{rot, boost} = \left( \frac{1}{1 - \frac{\varv}{\varv_{\rm crit}}} \right)^\xi
\end{equation}
where the exponent $\xi$ is 0.43. While there is ongoing debate on whether or not rotation enhances mass loss in canonical massive stars \citep{Muller14}, however when close to the break-up limit in near-Eddington situations, the formalism by \citet{Maeder00B} captures the effects of both radiation-driven wind mass loss and gravity-darkening effects. We therefore use the rotation boost factor by 
\citet{Maeder00B} in our rapid rotating massive star models close to their Eddington limit.

\subsubsection{Mechanical Mass Loss at break-up}
\label{sss-Methods_MechMdot}

\begin{figure}
    \centering
    \includegraphics[width=1.05\linewidth]{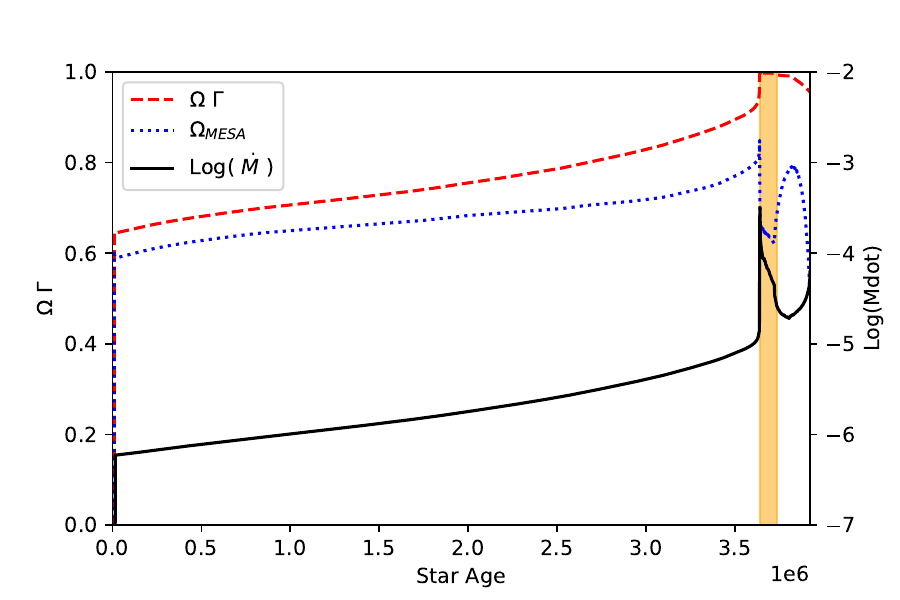}
    \caption{Evolution of a $100 M_{\odot}$ star with an initial rotation of $\Omega / \Omega_{\rm crit} = 0.6$ to the end of He burning. The mass-loss rate of the model is shown with the solid black line, while the evolution of rotation ($\Omega$) is shown with the dotted blue line and the $\Omega \Gamma$ value of Equation \ref{Eq-GammaOmega} is shown with the red dashed line. The orange highlighted region is where our condition for enhanced mass loss is satisfied, and the model is undergoing mechanical mass loss. Note that the value of $\Omega$ here is based on the MESA definition, which is different than the definition of \citet{Maeder00B}.}
    \label{Fig-100M_NoST_OG_O_M_plot}
\end{figure}

Finally the last component of our mass loss is called the mechanical mass loss. While the rotation boost factor from \citet{Maeder00B} increases as one approaches the $\Omega \Gamma$ limit, it is not sufficient to prevent models from crossing this limit. Especially when the models undergo a rapid total contraction phase following core H exhaustion. We evaluate the ratio of surface equatorial velocity to the critical velocity, either the first or the second depending on the value of the Eddington parameter, and if the ratio crosses $0.98$, we switch on a mechanical mass loss which gradually increases the mass-loss rate until the model safely returns below $\varv/\varv_\mathrm{crit,MM} = 0.98$. This can be switched on and off multiple times during the evolution whenever the $\varv/\varv_\mathrm{crit,MM}$ ratio goes beyond 0.98. The implementation is similar to that of \citet{Meynet06}. We chose a threshold of 0.98 instead of strict unity for numerical stability.

Now that we have discussed the different components to our mass loss recipe, we can write the overall mass loss as follows
\begin{equation}
    \label{Eq-MM00_MdotBoost_Generic}
    \dot{M}_\mathrm{tot} = f_\mathrm{rot,boost,MM} \times \dot{M}_\mathrm{rad}(\Omega = 0) + \dot{M}_\mathrm{mech}
\end{equation}
where the second term on the right side is only switched on when $\varv/\varv_\mathrm{crit,MM} = 0.98$ goes beyond 0.98. 

Figure \ref{Fig-100M_NoST_OG_O_M_plot} shows the evolution of a $100 M_{\odot}$ model (the same model as in Figure \ref{Fig-100M_06Om_Vcrit_plot}) highlighting where mechanical mass loss begins and ends. Once the $\varv/\varv_\mathrm{crit,MM} = 0.98$ condition is satisfied during the total contract phase following H exhaustion at roughly $\sim 3.6$ Myr, the mechanical mass-loss rate switches on. The absolute rate sharply shoots up to log$(\dot M ) \sim -3.5$ and drops from there as the star begins to stabilise. After a short period of this, $\varv/\varv_\mathrm{crit,MM} = 0.98$ drops below 0.98 and the mechanical mass loss switches off.

\begin{figure}
    \centering
    \includegraphics[width=1.05\linewidth]{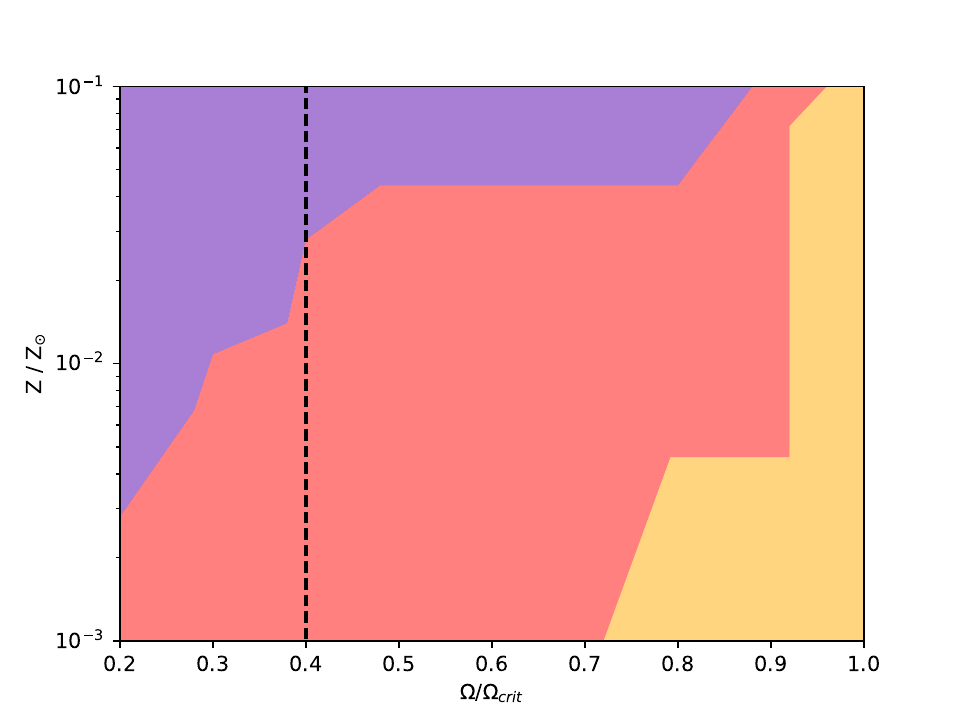}
    \caption{Contour plot showing which models become supercritical for $M_{\rm ZAMS} = 50 M_{\odot}$ across $\Omega / \Omega_{\rm crit}$ and $Z / Z_{\odot}$. Purple is for models which never reach supercriticality during central H or He burn, while red is for models which become supercritical at some point in their evolution. Yellow are models which are supercritical on the ZAMS. The black dashed line denotes the limit of our \citet{Winch24} rotation rates.}
    \label{Fig-50M_om_vs_Z_contourf}
\end{figure}

In Figure \ref{Fig-50M_om_vs_Z_contourf}, we show the parameter space in $Z$ and initial rotation, and where the different components of our mass loss recipe is being used. For slow rotators, a model can be very low ($\sim 1/1000$th $Z_\odot$) before reaching supercriticality, but this rapidly increases to $1/100$th $Z_\odot$ at $\Omega = 0.3 \Omega_{\rm crit}$. After this, however, the radiative mass loss is enough through the star's core H and Helium (He) burning lifetimes to not become supercritical during its evolution. However, at $\Omega \approx 0.75 \Omega_{\rm crit}$, the models start to become supercritical on the Zero Age Main Sequence (ZAMS) when at low Z (denoted by the yellow region). This naturally asymptotes to all metallicities as rotation is further increased to $\Omega \sim \Omega_{\rm crit}$.

\subsection{Testing pair instability and the $M_{\rm crit}$ experiment}
\label{ss-Methods_PI_MCrit}

Similar to \cite{Winch24}, we use a two-pronged approach here to investigate the BH/PPI boundary. We first run a small number of models forming a dedicated grid to investigate the location of the PI gap using the critical core mass as a criterion - the $M_{\rm crit}$ experiment. These models are run from their ZAMS to the end of core Oxygen burning - unless the model becomes unstable whereupon it stops early. Following this, our main grid is run from ZAMS to the end of core He burning, as at this point we can use the critical core mass determined from the $M_{\rm crit}$ experiment to determine which models become unstable.

Modelling the full PPISN or PISN scenario would require the use of MESA's HLLC hydrodynamics solver, and so instead we focus our investigation on the lead up to pair instability at the lower limit where the core-collapse scenario gives way to the PPISN regime. To determine when a model would become unstable, we use the first adiabatic index, $\Gamma_1$ given in equation \ref{Eq-gam1}. However, as this is a local value for the adiabatic index, we use the criterion of \citet{Stothers99} to integrate over the entire star and achieve a global condition for the stability of the star. This is given in equation \ref{Eq-int_gam1}, and is the basis of our $M_{\rm crit}$ experiment.

\begin{equation}
    \label{Eq-gam1}
    \Gamma_1 = \left( \frac{d \ln P}{d \ln \rho} \right)_{\rm ad}
\end{equation}

\begin{equation}
\label{Eq-int_gam1}
    \left< \Gamma_1 \right> = \frac{\int \Gamma_1 P d^3 r}{\int P d^3 r}
\end{equation}

This criterion can be used to derive a ``critical core mass'' which we document in Section \ref{ss-McritExperiment}. This critical core mass forms the boundary between the direct collapse scenario and PPI.

For our critical core mass, we are interested in the mass of both the He and Carbon-Oxygen cores in the final stages of the star's life. For the He core mass, we define its location as the innermost boundary where the abundance fraction of H is below 0.01, and where the abundance fraction of He is above 0.01. For the CO core, we define this as the location where the abundance of Carbon-Oxygen exceeds that of He. This is slightly different to other work in the literature, which uses the typical approach requiring the mass fraction of He to be below 0.01 as well. However, as we demonstrate in Section \ref{ss-McritExperiment}, a stripped star may have a small amount of He near the surface which remains due to the shrinking of the convective core during the last stages of core He burning. The uppermost layers of the star are no longer fully convective, which means that the He remains unmixed. Regardless, these regions are still almost completely Carbon-Oxygen, which leads us to use our definition of the CO core mass which would otherwise provide a conservative value if using the traditional method. 

Examining our main grid of models and the relationships within, we start by assuming the following relation;

\begin{equation}
    \label{Eq-Mbh_Methods}
      M_{\text{final}}~= ~M_{\text{core}} ~+ ~M_{\text{envl}} 
\end{equation}

From this, we expect that $M_{\rm core}$ and $M_{\rm final}$ are both a function of the metallicity, initial mass and rotation. This is different to our previous work, as this means that after a certain point $M_{\rm core} \approx M_{\rm final}$ due to the star becoming stripped, and thus $M_{\rm envl} = 0$. We use Equation \ref{Eq-Mbh_Methods} in Section \ref{ss-Disc_MCrit} to describe our relations for our fits around the PI boundary.

For our main grid, the input parameters are listed in Table \ref{T-Params}.

\begin{table}
\centering
\begin{tabular}{c c c c c}
\hline
$ { M_\mathrm{ZAMS} }$ & $Z/Z_{\odot}$ & $\alpha_{\text{ov}}$ & $\Omega/\Omega_{\text{crit}}$ & $\alpha_{\text{sc}}$ \\
\hline
\hline
50 &  1/10 & 0.1 & 0.4 & 1 \\
60 & 1/100 &  & 0.6 & \\
70 & 1/1000 &  & 0.8 &  \\
80 &  &  &  &  \\
90 &  &  &  &  \\
100 &  &  &  &  \\
\hline
\end{tabular}
\caption{List of physics parameters used in our model grid. We provide the initial mass (\Mi), metallicity ( $Z / Z_\odot$ ), over shooting parameter (\aOv), rotation rate as a function of critical rotation (\OmOmC), and semiconvective efficiency parameter (\aSC).}
\label{T-Params}
\end{table}

\section{Results}
\label{s-Results}

The effect of fast rotation for H-rich massive stars on the location of the PI boundary and the population of BH/PI progenitors is yet to be explored in stellar evolution. Recent work by \citet{Volpato24} has examined rotating low metallicity massive stars, however, this was not completed in a systematic way or with a focus on the BH-PPI boundary. 
In \cite{Winch24}, we found a constant critical core mass for the BH-PPI boundary for different stellar evolution parameters, and examined how these physical mechanisms can change the BH/PI progenitor properties of massive stars for a range of assumptions. However we were limited to comparatively moderate rotation rates of $\Omega \leq 0.4 \Omega_{\rm crit}$ - or even $\Omega \leq 0.2 \Omega_{\rm crit}$ for $Z = 1/1000$th $Z_{\odot}$. With our mechanical mass loss implementation, we can extend our investigation up to $\Omega = 0.8 \Omega_{\rm crit}$ and also provide new results in MESA for fast rotating, low metallicity stars at the PI boundary.

\subsection{Effects of Rotation on Models}
\label{ss-Results-EffectsOfRotation}

At the end of the main sequence, there is a significant numerical challenge in evolving rapidly rotating models into core He burning. The rapid increase in luminosity, further exacerbated by a high \OmOmC, leads to a very rapid evolution towards the $\Omega \Gamma$-limit. Our method for resolving this is addressed in Section \ref{sss-Methods_MechMdot}. 

In this subsection, our aim is to test what effect our rotation implementation has on our models with a small subset of models which we can compare since this is the first time the mechanical mass loss formulae of \citet{Meynet06} have been implemented in MESA to our knowledge. Table \ref{T-H_rich_MechMdotBreakdown} \textcolor{black}{documents the models we use for this subsection}, and the relevant results to this discussion. The main conclusions we draw from this table are as follows;

\begin{figure}
    \centering
    \includegraphics[width=1.05\linewidth]{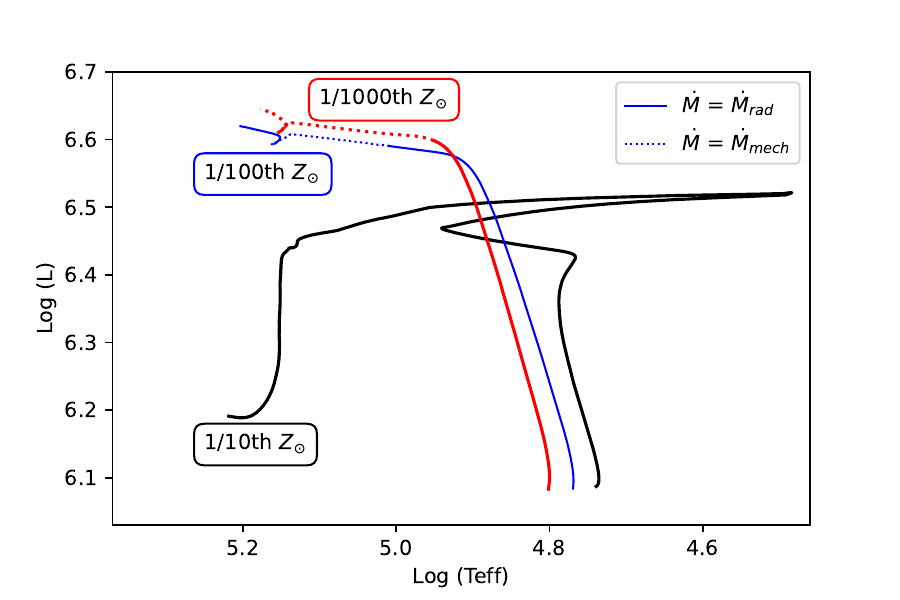}
    \caption{Hertsprung-Russel Diagrams of $100 M_{\odot}$ models from $1/10$th $Z_{\odot}$ to $1/1000$th $Z_{\odot}$ \textcolor{black}{with an initial rotation rate of $\Omega = 0.6 \Omega_{\rm crit}$ (as per models B1, B2, B3 in Table \ref{T-H_rich_MechMdotBreakdown})} from the ZAMS until end of core He burning. Regions of mechanical mass loss are shown using a dotted line in place of a solid line, which represents typical evolution using only radiative wind mass loss. All models are without a Spruit-Tayler Dynamo.}
    \label{Fig-100M_HRDs_NoST_MechMdot}
\end{figure}

\begin{figure}
    \centering
    \includegraphics[width=1.05\linewidth]{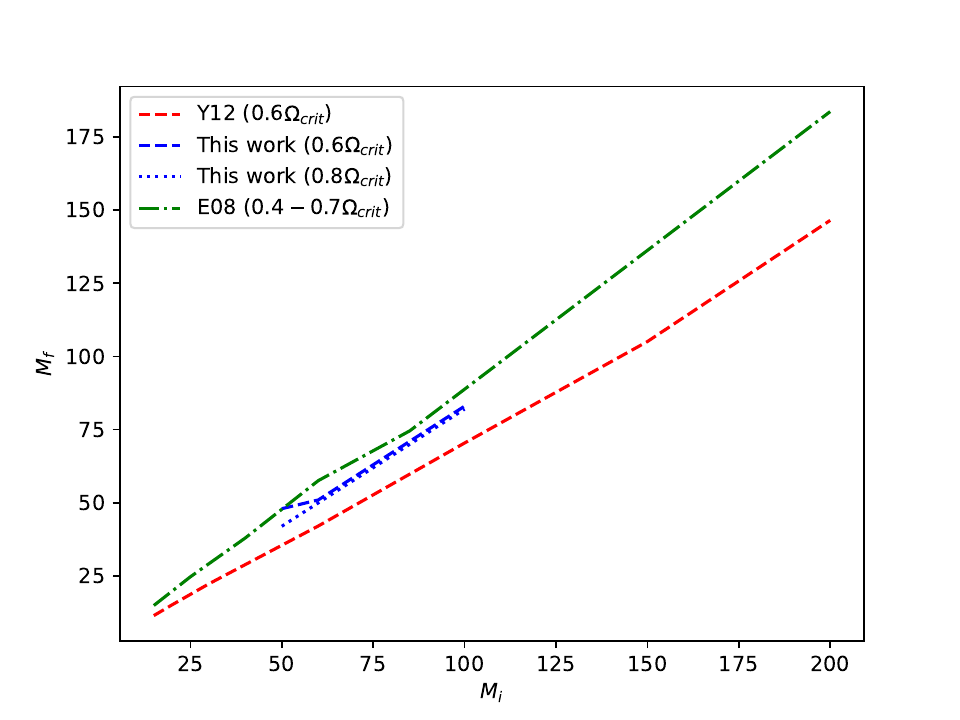}
    \caption{Final mass vs initial mass for the works of \citet{Ekstrom08, Yoon12} (E08, and Y12 in the legend respectively) in comparison to this work. The green dot-dashed line follows the final vs initial mass relationship in \citet{Ekstrom08} while the red dashed line traces the results of \citet{Yoon12}. Our results are in blue for models at both $0.6$ and $0.8 \Omega_{\rm crit}$ \textcolor{black}{for $Z = 1/1000$th $Z_\odot$} using models from the main grid in Section \ref{ss-MainGrid} in order to encompass the differences in rotation rate (and the assumptions therein) between models in either paper.}
    \label{Fig-Y12_E08_MisComparison}
\end{figure}

\begin{itemize}
    \item Models B1, B2 and B3 (tracks for these are shown in Figure \ref{Fig-100M_HRDs_NoST_MechMdot}) have the same initial mass and $\Omega / \Omega_{\rm crit}$ for a range of metallicities, thus isolating the effect of decreasing metallicity on the amount of mass loss - specifically the contribution through mechanical mass loss. While the highest total mass loss is in model B1 ($Z = 1/10$th $Z_\odot$), this model has no mechanical mass loss as the radiative winds are strong enough to avoid critical surface rotation rates. However, decreasing metallicity one notices a trend in the proportion of mechanical mass loss as a fraction of the total (column 8). Lower metallicity may result in less mass loss overall, but much more of a contribution from mechanical mass loss - from $35 \%$ of the total for model B2, to $70 \%$ of the total for B3. \\
    \item Comparing models A1 and B2 presents us with information regarding the mass dependence of this relationships. This time, the metallicity is kept constant between these models as is $\Omega / \Omega_{\rm crit}$, however these two models have different initial masses. In this comparison we are interested in the differences in the mechanical mass loss as a fraction of the total (column 8) \textcolor{black}{of Table \ref{T-H_rich_MechMdotBreakdown}}. It is interesting to note that the fraction of mechanical mass loss is higher for model A1, at $M_{\rm ZAMS} = 60 M_\odot$, than for model B1, at $M_{\rm ZAMS} = 100 M_\odot$. This is because at $100 M_\odot$, the strength of the radiative winds is higher due to much higher luminosities, and thus there is less angular momentum to strip by the time the star encounters mechanical mass loss. \\
    \item Models with a Spruit-Tayler Dynamo (marked by the suffix ``ST'' on their model name) have several behaviours which are interesting to draw out here. Firstly, examining models B1 and B1ST ($Z = 1/10$th $Z_\odot$), we notice that here is where the presence of ST has the most effect on the final mass of the model (shown in the final column) with a difference between the model's final mass of over $10 M_\odot$. This significance is not true at lower metallicity. For models B2ST and B3ST ($Z = 1/100$th, $1/1000$th $Z_\odot$ respectively) there is almost no difference between their non-magnetic counterparts (B2 and B3) in the final mass with barely $1M_\odot$ difference between the magnetic and non-magnetic models. This is in direct opposition to the stipulation by \citet{Yoon12} who suggests that the key difference between their models and those run using the Geneva code in \citet{Ekstrom08} is the presence or lack thereof of the ST Dynamo in the calculation of the evolutionary models. We shall go into detail on the exact causes of these differences later in this section to show that this is not the case. \\
    \item While differences in rotational mixing implementation and mass-loss rates do create uncertainties in the final mass as seen in Figure \ref{Fig-Y12_E08_MisComparison}, this does not translate into uncertainties in $M_{\rm crit}$. This is because the underlying physics behind the $M_{\rm crit}$ core mass does not change with rotation or mass loss and is only dependent on the pair production within the core.
\end{itemize}

\begin{table*}
\centering
\begin{tabular}{c || c c c | c c c c | c}
\hline
Model & $ { M_\mathrm{ZAMS} }$ & $Z/Z_{\odot}$ & $\Omega/\Omega_{\text{crit}}$ & $\Delta \dot M_{\rm Total}$ & $ \Delta \dot M_{\rm Mech}$ & $ \Delta \dot M_{\rm Rad}$ & $\frac{\Delta M_{\rm Mech}}{\Delta M_{\rm Total}}$ & $M_{\rm final}$\\
\hline
\hline
A1 & 60  & 1/100  & 0.6 & 9.72  & 7.74  & 1.98  & 0.79 & 50.28 \\
B1 & 100 & 1/10   & 0.6 & 63.36 & 0     & 63.36 & 0    & 36.64 \\
B2 & 100 & 1/100  & 0.6 & 20.66 & 7.22  & 13.44 & 0.35 & 79.34 \\
B3 & 100 & 1/1000 & 0.6 & 17.07 & 12.02 & 5.05  & 0.70 & 82.93 \\
\hline
A1ST & 60  & 1/100  & 0.6 & 9.75  & 7.75  & 2.00  & 0.79 & 50.25 \\
B1ST & 100 & 1/10   & 0.6 & 51.00 & 0     & 51.00 & 0    & 49.00 \\
B2ST & 100 & 1/100  & 0.6 & 21.64 & 6.58  & 15.06 & 0.30 & 78.36 \\
B3ST & 100 & 1/1000 & 0.6 & 17.77 & 13.45 & 4.32  & 0.76 & 82.23 \\
\hline
\end{tabular}
\caption{Breakdown of total and mechanical mass loss for several H rich models with across a range of metallicities. Models A1 through B3 do not have an ST dynamo, while A1ST through B3ST use an ST dynamo with the efficiency parameter $D_{\rm ST} = 1$. $\Delta M_{\rm Total} = \Delta M_{\rm Rad}\cdot f_{\rm rot, boost, MM} + \Delta M_{\rm Mech}$ as per Section \ref{s-Methods}.}
\label{T-H_rich_MechMdotBreakdown}
\end{table*}

There is an ongoing debate about whether or not to use the ST dynamo and how significant this choice is. The Bonn models typically employ the ST dynamo \citep[][]{Yoon12}, while Geneva models do not \citep[][]{Ekstrom08}. While this will certainly cause some difference in the evolution of these massive star models, we see that for the low metallicity case these differences are negligible if just considering whether or not the ST dynamo is switched on or off. We notice a significant difference in ST or no-ST models at higher metallicities as seen with models B1 and B1ST, however the differences in evolution and specifically the final mass of rotating (very) massive star models between Bonn/MESA (using diffusive mixing only) and Geneva (using advective-diffusive mixing) is expected to be the result of the choice of mixing implementation in the evolution code and the mass-loss prescription used, rather than the inclusion/exclusion of the ST dynamo based on our results. We present a comparison between the final mass and initial mass of \citet{Ekstrom08, Yoon12}, and our models in Figure \ref{Fig-Y12_E08_MisComparison}. Comparing to similar models in both \citet{Yoon12, Ekstrom08}, we see that our models lie in between both of the expected values in those papers. To take a specific example, the $60 M_\odot$ model in \citet{Ekstrom08} finishes its evolution with  $58 M_\odot$ meanwhile our models predict a final mass of $\sim 50 M_\odot$ for a similar metallicity and rotation rate. The comparative model in \citet{Yoon12} finishes with a final mass of $42 M_\odot$. 

The difference between the $60 M_\odot$ model from \citet{Ekstrom08} and ours is caused by the result of the different approaches to rotational mixing between MESA and Geneva. Geneva models cool down in surface temperature with high rotation, while MESA models tend towards hotter temperatures. This causes our model to evolve with more mass loss over time. This difference was verified by running a $60 M_\odot$ model in MESA by removing mixing parameter for Eddington-Sweet circulation ($D_{\rm ES} = 0$). This does not make MESA replicate Geneva's advective-diffusive mixing implementation, but it has the effect on the HRD of making the model turn towards to the cool-end of the HRD. This model's final mass was $59 M_\odot$, which is almost exactly the same as the corresponding model in \citet{Ekstrom08}. 

Conversely one can see in the Bonn tracks from \citet{Yoon12} that their models also turn towards the blue end of their HRD (their Figure 1) when rotating. This leads to the mass loss difference when compared to the Geneva models. However, the extra mass loss compared to the models in this work is instead explained by the choice of radiative mass loss recipe and the choice of how to deal with the mechanical mass loss, specifically the choice of \citet{NugisLamers00} which has very high mass-loss rates compared to \citet{Sander20} \citep[which is included in][]{Sab23} for He stars. 

\subsection{$M_{\rm crit}$ Experiment at High Rotation}
\label{ss-McritExperiment}

\begin{figure*}
    \centering
    \includegraphics[width=1\linewidth]{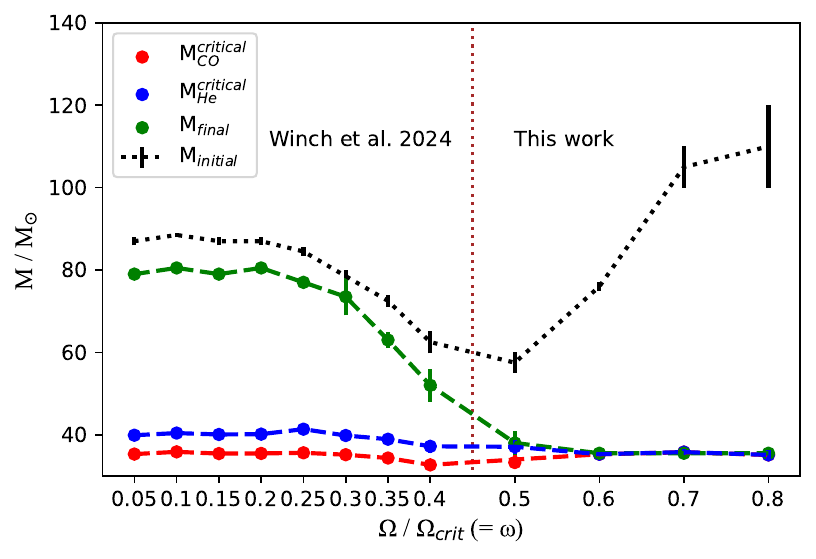}
    \caption{$M_{\rm crit}$ experiment extension from \citet{Winch24}. Data points indicate the $M_{\rm crit}$ values for that $\Omega / \Omega_{\rm crit}$, and error bars are the uncertainties from the models used. Models above $\sim 0.5 \Omega_{\rm crit}$ become fully mixed stripped stars and thus by the end of O burn, the star becomes a complete CO core. \textcolor{black}{For these models we use the fiducial metallicity of $Z = 1/10$th $Z_\odot$ to keep consistency with the original work.}}
    \label{Fig-MCrit_HighRotation}
\end{figure*}

With the effects of rotation and our mass-loss enhancement established in the previous subsection, here we extend the $M_{\rm crit}$ experiment up to $0.8 \Omega_{\rm crit}$, where mechanical mass loss plays a part. Note that for all remaining models in this paper, we turn off the Spruit-Tayler dynamo such that $D_{\rm ST} = 0$ as we show in Section \ref{ss-Results-EffectsOfRotation} that ST does not have a significant effect at low metallicity, and our models in \cite{Winch24} did not use the ST dynamo. Figure \ref{Fig-MCrit_HighRotation} shows both our first paper results for context, and also the new results from high rotation models. In \cite{Winch24}, we focused on the mass of the He core for our critical core mass, though we also provided a value for the Carbon-Oxygen critical core mass, and we created fits using the He core mass. However, for our fast rotators, the critical core mass for He drops to the same value as the Carbon-Oxygen core mass since these stars become stripped stars due to their rotation. This means that the He core mass is no longer constant as in \cite{Winch24} - this can be seen in Figure \ref{Fig-MCrit_HighRotation} as the blue line denoting the Helium core decreases and matches the red line denoting the CO core mass at $0.6 \Omega_{\rm crit}$. Thus, we focus on the value of CO critical core mass which remains constant for high and low rotation. The critical CO core mass for fast rotators ($\Omega > 0.4 \Omega_{\rm crit}$) is constant at $M_{\rm CO, crit, high \Omega / \Omega_{\rm crit}} = 35.0 \pm 2.3 M_{\odot}$. While this value is lower than the previous number from \cite{Winch24} due to the transition between H-rich and stripped star progenitors at \textcolor{black}{$0.4 - 0.5 \Omega_{\rm crit}$}, it is within the error bounds established in \cite{Winch24}.

We have seen in the previous section how mass loss and rotation are connected; high rotation leads to high mass loss, brakes the star, leads to lower rotation, either through rotationally enhanced mass loss or breakup mass loss. For a model \textcolor{black}{in our $M_{\rm crit}$ Experiment} at $0.8 \Omega_{\rm crit}$, with an initial mass of $M_{\rm ZAMS} = 120\,M_\odot$, the star becomes instantly almost completely chemically homogenous, then loses mass and angular momentum at rapid rates. By the end of core Oxygen burning, the star is rotating $\sim 3$ times slower than its $M_{\rm ZAMS} = 60\,M_\odot$, $0.5 \Omega_{\rm crit}$ counterpart. This is different to what is seen in the models for smaller $\Omega / \Omega_{\rm crit}$, where faster rotators will still finish with faster rotation as the mechanical mass loss of these models is far greater.

\begin{figure}
    \centering
    \includegraphics[width=1.05\linewidth]{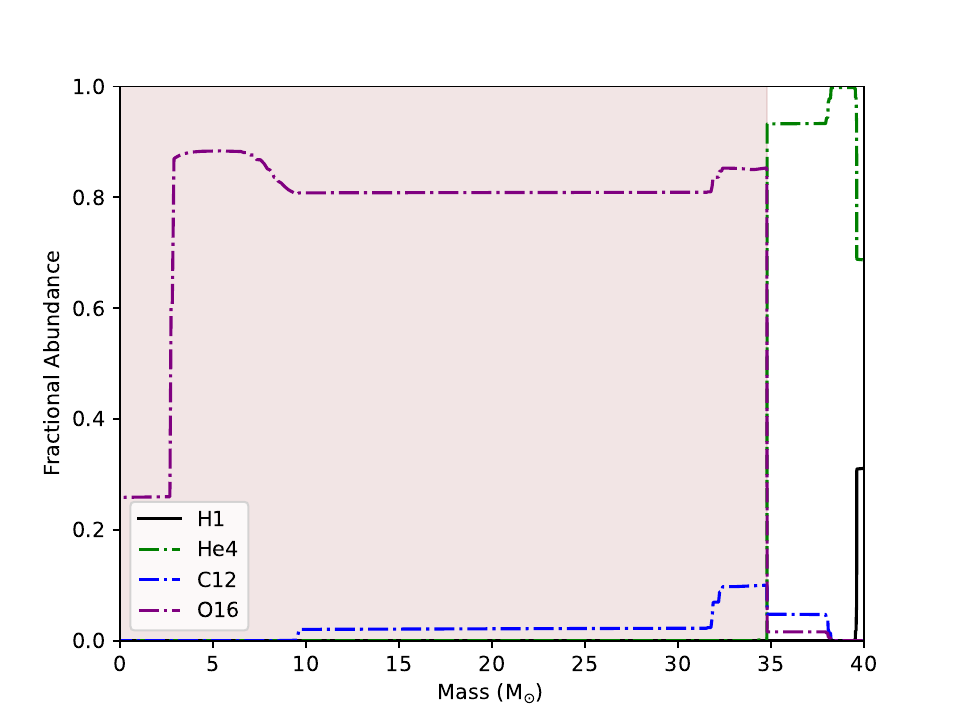}
    \label{Fig-HRich_Profile}
    \includegraphics[width=1.05\linewidth]{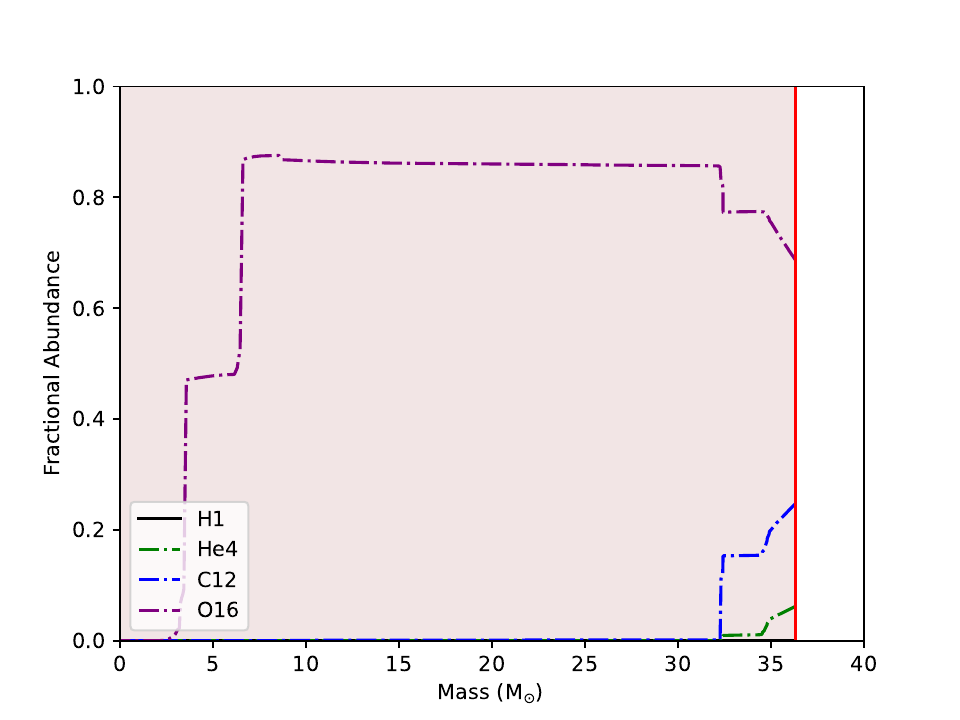}
    \caption{Profiles for two stars showing how the definition of the CO core can change the mass coordinate of the core. The top panel shows a H rich star with $M_{\rm ZAMS} = 78 M_{\odot}, \Omega = 0.3 \Omega_{\rm crit}$, while the bottom panel shows a stripped star with $M_{\rm ZAMS} = 120 M_{\odot}, \Omega = 0.8 \Omega_{\rm crit}$ both at the end of core Carbon burning. The dot-dashed lines represent abundances, while the shaded region is the CO core. The red line in the bottom panel is the stellar surface. Note that the surface of the H rich model's surface is beyond the limit of the plot.}
    \label{Fig-StrippedStarProfile}
\end{figure}

Now we aim to understand why our value of the $M_{\rm CO, crit}$ for high rotation is consistent with the value from \cite{Winch24}. One possible alternative is that the presence of the H-rich envelope creates a stabilising effect on the integral of $\Gamma_1$ as this is an integral taken over the entire star, and the envelope itself is dynamically stable. Thus, we now define two values of $\left< \Gamma_{1} \right>$; $\left< \Gamma_{\rm Total} \right>$ herein will be the average of the adiabatic index over the entire star (as we have used previously), while $\left< \Gamma_{\rm Core} \right>$ will be the average for only the He core - similar to the experiment in \citet{Costa21} and we will use the same criteria. By doing so, we find that for stripped stars, the difference between $\left< \Gamma_{\rm Total} \right>$ and $\left< \Gamma_{\rm Core} \right>$ is nearly zero, as expected. For stars which retain a significant H envelope, we find that while there are differences in $\left< \Gamma_{\rm Total} \right>$ and $\left< \Gamma_{\rm Core} \right>$, these values are minimal and almost certain to not have a significant effect on the determination of whether a model is pair unstable or not. This is in agreement with the results of \citet{Costa21}.

\textcolor{black}{An alternate explanation for the changes in $M_{\rm CO, crit}$ could be the presence, or absence, of a thin Helium shell above the CO core. }As discussed in Section \ref{s-Methods}, stripped stars may have a small amount of He near their surface which, depending on one's definition of the He-core and CO-core boundaries, could change the value of these core masses - and thus the value of the \textit{critical} core mass of our experiment - by $\sim 3 M_{\odot}$. However, these changes are only due to definitions, and not changes in the structure of the star. Figure \ref{Fig-StrippedStarProfile} shows two profiles of models in our high-rotation $M_{\rm crit}$ experiment grid. It can be seen that for the innermost $40 M_{\odot}$ of both stars (note that the top panel showing the H-rich star is only showing the innermost $40 M_\odot$ and the rest of the star is not shown) that the slow rotator ($0.3 \Omega_{\rm crit}$) has a sharp CO-He boundary whereas the fast rotating stripped star has a gradient of CO and He abundance near the surface. This is caused by the recession of the convective core in the final states of the star's core He burning period, where this material does not get mixed in with the rest of the core. This is only a very thin layer, but as the core recedes with the star's mass loss, this is sufficient to leave behind a gradient. As this is only a minor structural change and not one which changes the shape of $\Gamma_1$ significantly through the core (and thus changing the limit for PI), we remain with our definition of the CO core mass which shows consistency with our original work.

\subsection{High Rotation Model Grid}
\label{ss-MainGrid}

Using the results of the previous section, we now aim to complete our original parameter space in \cite{Winch24} by filling in the fast-rotators. We achieve this with a smaller grid of models, the parameter space of which is detailed in Table \ref{T-Params} \textcolor{black}{in Section \ref{s-Methods}}. 

\textcolor{black}{These models were ran from the ZAMS to the end of core Helium burning, and accompanies the models from Section 4 of \citet{Winch24}. The overall pattern in these models is described in detail in this paper in Section \ref{ss-Results-EffectsOfRotation}. However, as this is a much larger grid with systematic models across the parameter space, we are able to find fits for these models.}
Aside from a few outliers, rotation has the same general pattern as seen with metallicity and overshooting - that is to say that increasing rotation results in more mass loss with all else being independent, and this difference is greater for higher initial masses. 

We produce two fits for these rotating models based on the data. The first is the fit for the core mass which is described in Equation \ref{Eq-MCoreFit}. Due to the nature of rotation, at some point the entire star will evolve chemically homogenously regardless of other initial conditions. We thus use $M_{\rm final, fit}$ in the cases where the fit for $M_{\rm CO}$ would be larger than the final mass. Note that the reason why our fit for $M_{\rm CO}$ can become larger than $M_{\rm final, fit}$ is because rotation also increases the core mass. A fit for $M_{\rm CO}$ takes account of this, but does not account for the physical impossibility for a core mass to be larger than the total mass of the star. We account for this with the aforementioned conditions.

The relation between $M_{\rm CO}$ and $M_{\rm ZAMS}$ is approximately linear, with rotation having only a small change on the final core mass as, while rotation does act like overshooting in that it extends the upper boundary of the convective core, the mass of the core can only grow to the total mass of the star as an absolute maximum. Meanwhile different metallicities result in different mass-loss rates which changes how much the core mass can grow by. 

For the final mass, the relationship here is more complicated. While for each variable of $M_{\rm final, fit}$, the relationship is approximately linear, combining these variables produces a much more complex picture. As such we produced $M_{\rm final, fit}$ using a polynomial regression model to derive a relationship for the fit.

\begin{equation}
M_{\rm CO, fit} = 
\begin{cases} 
3.32 + 0.86 M_{\rm ZAMS} - 0.23 \frac{Z}{Z_\odot}^{0.44} M_{\rm ZAMS}^{0.98} & M_{\rm CO} < M_{\rm final} \\
M_{\rm final, fit} & M_{\rm CO} \geq M_{\rm final}
\end{cases}
\label{Eq-MCoreFit}
\end{equation}

The corresponding fit for $M_{\rm final, fit}$ is below.

\begin{equation}
\begin{split}
    M_{\rm final, fit} = M_{\rm ZAMS}\left( 1.09 - 0.6\Omega / \Omega_{\rm crit} + 0.0007M_{\rm ZAMS} - 2.98 \frac{Z}{Z_\odot} \right) \\ {
    - \Omega / \Omega_{\rm crit} \left( 11.11 - 36.81\Omega / \Omega_{\rm crit} + 380 \frac{Z}{Z_\odot} \right) + \frac{Z}{Z_\odot} \left( 190 + 177 \frac{Z}{Z_\odot} \right)}
    \label{Eq-MfinalFit}
\end{split}
\end{equation}

These fits were also produced with some of the data from \cite{Winch24} to ensure consistency in the fitting parameters. The error on the low-rotation $M_{\rm CO, fit}$ is $\pm 9.09 M_\odot$, while the error on $M_{\rm final, fit}$ is $\pm 2.86 M_\odot$.

\begin{figure}
    \centering
    \includegraphics[width=1.05\linewidth]{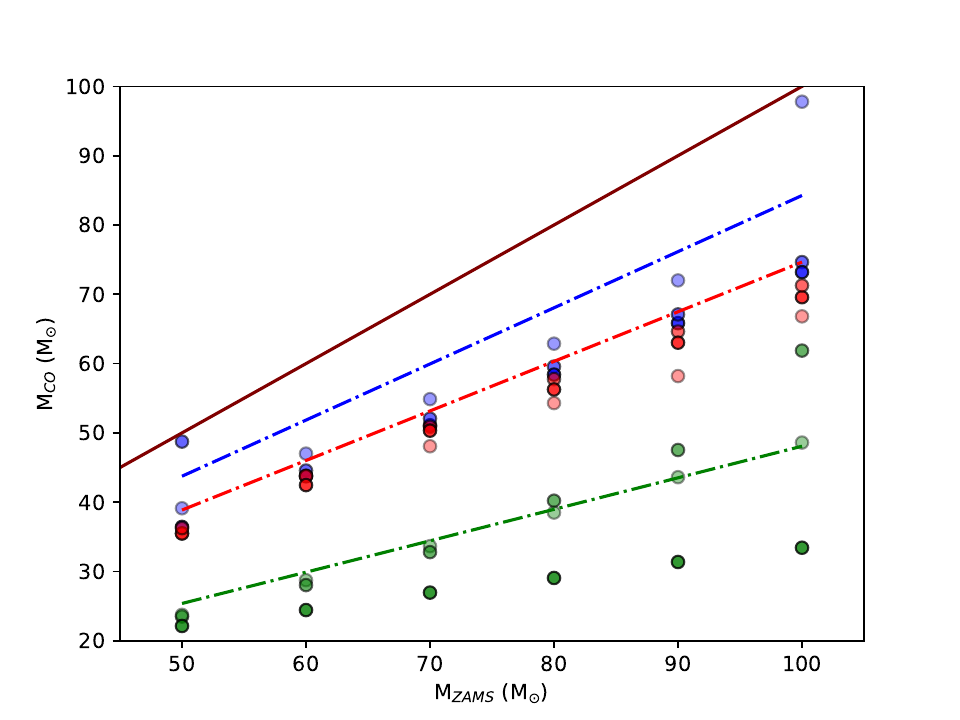}
    \caption{Combined plot of the datapoints for the main grid including fit lines for $M_{\text{CO} }$. The solid maroon line denotes the 1:1 line for mass such that any deviation from this line represents mass loss. Blue, red and green data points are for models with metallicities of 1/1000th, 1/100th and 1/10th $Z_\odot$ respectively. Models are shaded according to their rotation rates \textcolor{black}{- the lightest, medium and darkest shades correspond to $\Omega / \Omega_{\rm crit} = 0.4, 0.6, 0.8$ respectively}. The dashed line is for our fits for $M_{\text{CO} }$.}
    \label{Fig-MCcore_MZams}
\end{figure}

\begin{figure}
    \centering
    \includegraphics[width=1.05\linewidth]{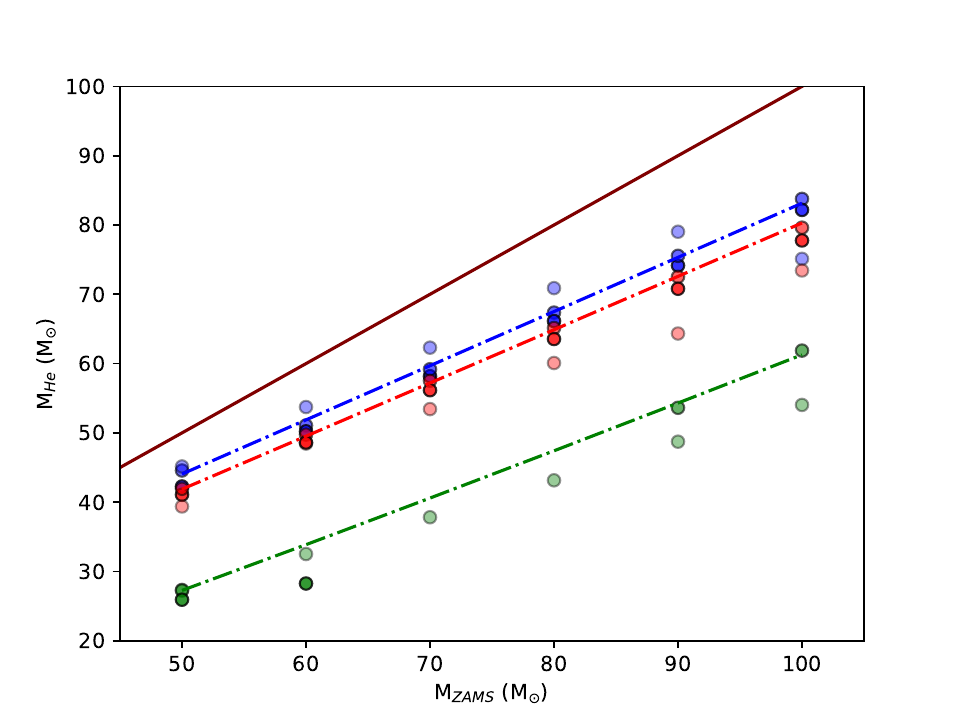}
    \caption{Combined plot of the datapoints for the main grid including fit lines for $M_{\text{He} }$. The solid maroon line denotes the 1:1 line for mass such that any deviation from this line represents mass loss. Blue, red and green data points are for models with metallicities of 1/1000th, 1/100th and 1/10th $Z_\odot$ respectively. Models are shaded according to their rotation rates \textcolor{black}{- the lightest, medium and darkest shades correspond to $\Omega / \Omega_{\rm crit} = 0.4, 0.6, 0.8$ respectively}. The dashed line is for our fits for $M_{\text{He} }$.}
    \label{Fig-MHecore_MZams}
\end{figure}

\begin{figure}
    \centering
    \includegraphics[width=1.05\linewidth]{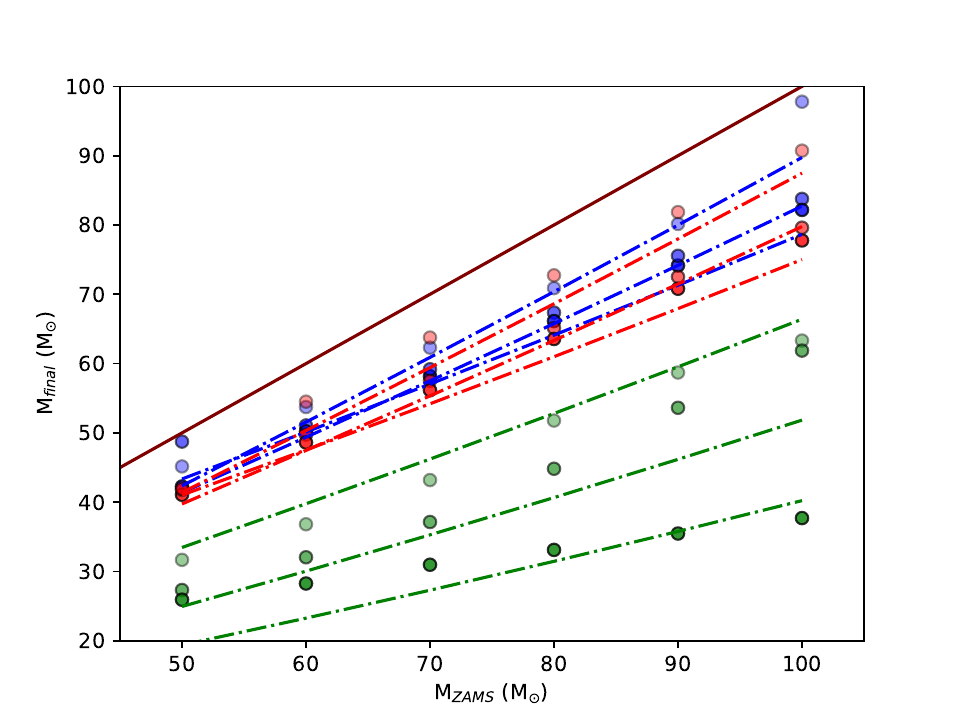}
    \caption{Combined plot of the datapoints for the main grid including fit lines for $M_{\text{final} }$. The solid maroon line denotes the 1:1 line for mass such that any deviation from this line represents mass loss. Blue, red and green data points are for models with metallicities of 1/1000th, 1/100th and 1/10th $Z_\odot$ respectively. Models are shaded according to their rotation rates \textcolor{black}{- the lightest, medium and darkest shades correspond to $\Omega / \Omega_{\rm crit} = 0.4, 0.6, 0.8$ respectively}. The dashed line is for our fits for $M_{\text{final} }$.}
    \label{Fig-Mfinal_MZams_fit}
\end{figure}

While the CO core fit provides a better approximation, we also provide a He core fit for consistency with our previous work. The He core fit is as follows:

\begin{equation}
M_{\rm He, fit} = 
\begin{cases} 
5.25 + 0.78 M_{\rm ZAMS} - 25.74 \frac{Z}{Z_{\odot}}^{0.82} M_{\rm ZAMS}^{0.38} & M_{\rm He} < M_{\rm final} \\
M_{\rm final, fit} & M_{\rm He} \geq M_{\rm final}
\end{cases}
\label{Eq-MCoreHeFit}
\end{equation}

while the root mean square error of $M_{\rm He, fit}$ is $\pm 3.04$.

The fit itself is still accurate to the data, despite not including rotation, as for these models the core mass increase from rotation is minor compared to the core mass differences due to metallicity, especially for low metallicity models. Furthermore, due to nonlinear effects, there are some cases where slower rotation may end with a larger core - especially at the high mass/high mass loss regime. At higher metallicities ($Z \geq 1/10$th $Z_\odot$), the deviations caused by rotation become more obvious as seen in Figure \ref{Fig-MCcore_MZams}. For the calculation of PI progenitors, this is not  a significant concern as the majority of these high mass models will still be above the PI limit. Figure \ref{Fig-MHecore_MZams} presents these models and the fit for them. Models which did not have a He core at the end of their evolution are excluded. Finally, Figure \ref{Fig-Mfinal_MZams_fit} presents our $M_{\rm final}$ fit versus the datapoints in a similar graph.

\subsection{Population Synthesis}
\label{ss-result-popsynth}

\begin{figure*}
\begin{subfigure}{\textwidth}
  \centering
  \includegraphics[width=0.8\linewidth]{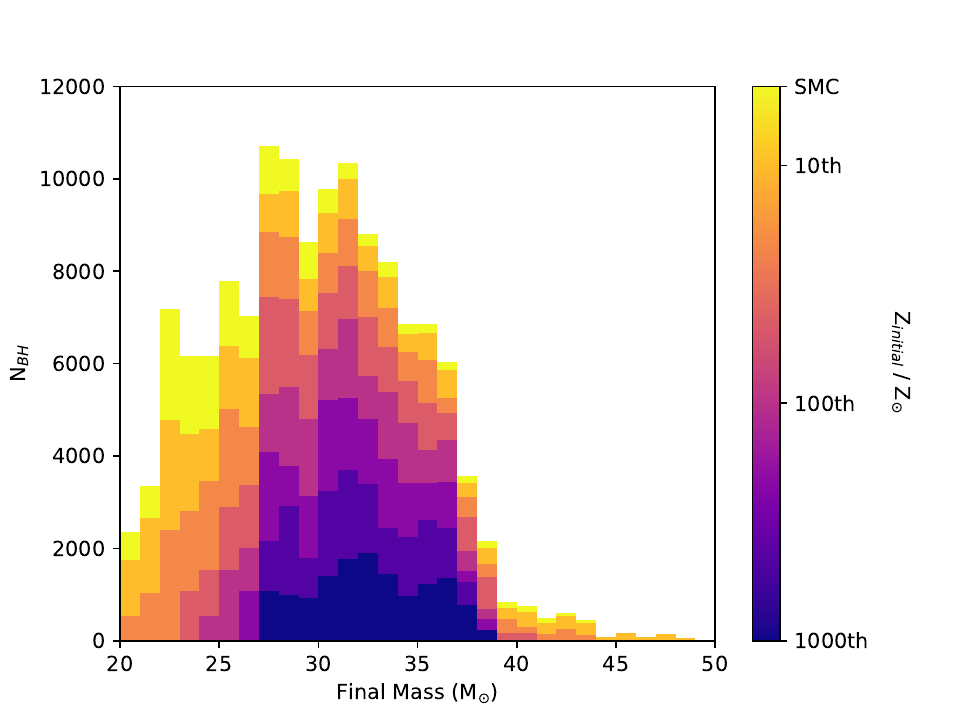}
  \caption{}
\label{Fig-BH_dist_Z}
\end{subfigure}
\begin{subfigure}{\textwidth}
  \centering
  \includegraphics[width=0.8\linewidth]{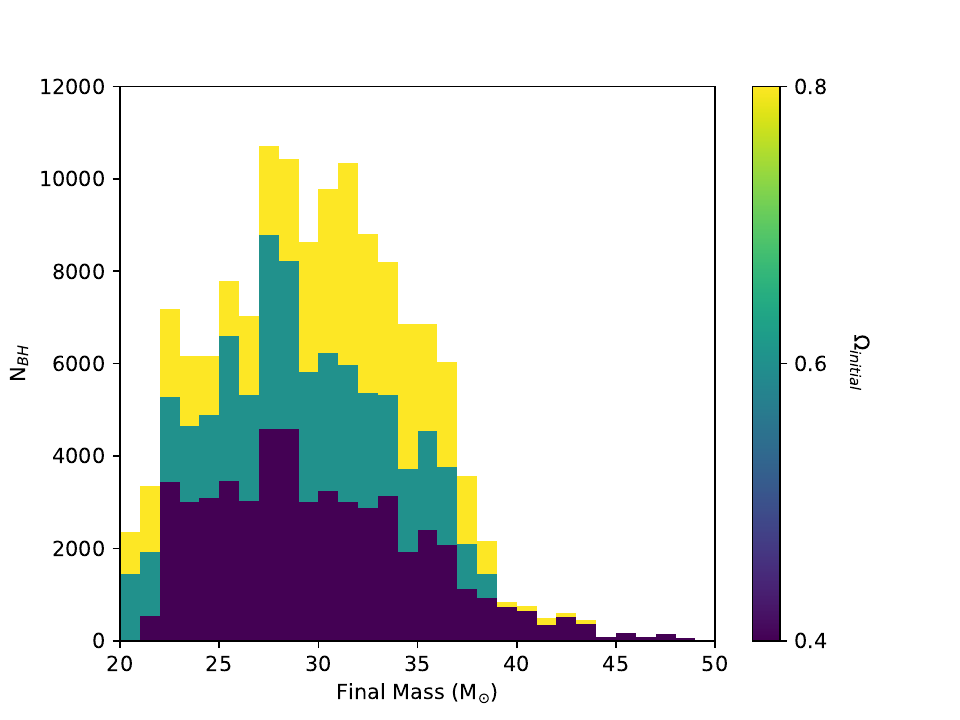}
  \caption{}
\label{Fig-BH_dist_om}
\end{subfigure}
\caption{Graphical representation of the numbers plotted in Table \ref{T-IMF_Table_BH}. Colours are arranged based on logarithmic metallicity in subfigure \ref{Fig-BH_dist_Z}, and $\Omega / \Omega_{\rm crit}$ in subfigure \ref{Fig-BH_dist_om}. The x-axis represents each mass bin, while the y-axis is the total number of stars in that bin, coloured based on each parameter.}
\label{Fig-BH_dist_bars}
\end{figure*}

Similarly to \cite{Winch24}, we demonstrate the use of our fits by producing a population of black hole progenitors in a simple population synthesis. We again choose a population of 240,000 stars for 3 different values of rotation ($\Omega / \Omega_{\rm crit} = 0.4, 0.6, 0.8$ weighed equally) and across 8 different metallicities (logarithmic through $Z / Z_{\odot} = 1/5  - 1/1000$th and also weighted equally). \textcolor{black}{We choose the initial mass function (IMF) of \citet{Salpeter55} ($M_{-2.35}$) to maintain consistency with our previous work. This is then weighed against our 240,000 star sample}. We apply our CO critical core mass criterion from Section \ref{ss-McritExperiment} into our fit from Equation \ref{Eq-MCoreFit}, and final mass fit from Equation \ref{Eq-MfinalFit}. Figure \ref{Fig-BH_dist_bars} provides a distribution of these models separated by rotation rate while Table \ref{T-IMF_Table_BH} provides the same data in tabular form. 

There are several striking differences between the bar charts in Figure \ref{Fig-BH_dist_bars} and that of our previous work for low rotation \citep[Fig. 10,][]{Winch24}. The first is the much lower maximum predicted BH mass, which is $\sim 50 M_\odot$ assuming there is still some shell left on top of the CO core, or $\sim 36 M_\odot$ if the star has been completely mixed. Note that this BH mass prediction does not include detailed simulation of the core collapse and thus the actual BH mass may be lower due to excess angular momentum during the infall. However, regardless of the final moments of the star during this infall, the mass loss history of these stars is already drastically different and our fits predict a robust upper limit to the mass of these black holes pre-PPI. However that is not the exclusive reason, as these stars are stripped stars, meaning that their core mass is their final mass in most cases. As such their final mass is then limited to the critical core mass as discussed above. The second notable feature is the large population feature in the $27-36 M_\odot$ range and the subsequent drop above $36M_\odot$. This is due to the aforementioned critical core mass criterion which limits these stars to a maximum total mass of $\sim 36 M_{\odot}$ if $M_{\rm final} = M_{\rm core}$. However, there are still stars above this which have a small He layer not mixed in to the CO core.

Another feature to note in the plots is the distribution of each of the rotation rates (subfigure \ref{Fig-BH_dist_om}), and the ratio of each rotation rate to the other within specific mass bins. Whereas in \cite{Winch24}, moving to higher final mass bins reduced the number of free parameters allowed to produce BH progenitors in those bins, the distribution in this study is mostly equal for each bin - specifically in the $27-36 M_\odot$ range. Thus having a higher rotation does not bias a model to any particular mass in this range as the cores are already large, even for $0.4 \Omega_{\rm crit}$. This is untrue after the critical core mass limit at $\sim 36 M_\odot$, however, as only models with $0.4 \Omega_{\rm crit}$ are left with a He shell by the end of core He burning in these fits. 

A very interesting feature we notice in our population study that appears to agree with the location of an observed bump in black holes in GWTC-3 \citep[Figure 10, ][]{Abbott21c} centred at $\approx 35 M_\odot$. As we see in our population, Figure \ref{Fig-BH_dist_bars}, we also have the majority of our population coalesce at a similar point. Based on our results in Section \ref{ss-result-popsynth}, we conclude that what we see with the distribution of stripped stars could be a contributor to this bump. As we mention, black holes from stripped stars are limited to their core masses due to the critical core criterion, and their masses at low metallicity are significantly reduced by strong mechanical mass loss. Thus, this may be a point for further study and discussion when synthesising the progenitor population of observed GW events. 

Finally, the tail of the distribution is made from stars which are both high metallicity and slow rotators. This is because, as mentioned, these stars need to not be fully mixed in order to go above the critical core mass limit. For this to be the case, the star would need to be a relatively slow rotator ($0.4 \Omega / \Omega_{\rm crit}$ in our parameter space) to keep the core as small as possible as a fraction of the total mass. Additionally, the model would have to have a relatively high metallicity ($> 1/100$th $Z_\odot$ and above) to not experience mechanical mass loss and thus keep an appreciable amount of mass above the CO core.

\begin{table*}
    \centering
    \begin{tabular}{ c | c | c c c | c c c }
    \hline
       $M_{\text{BH}}$ & $N_{\text{BH}}$ & $N_{\text{BH } \Omega = 0.4 \Omega_{\rm crit}}$ & $N_{\text{BH } \Omega = 0.6 \Omega_{\rm crit}}$ & $N_{\text{BH } \Omega = 0.8 \Omega_{\rm crit}}$ & $N_{\text{BH SMC-10th}}$ & $N_{\text{BH 10th-100th}}$ & $N_{\text{BH 100th-1000th}}$ \\
       \hline
       \hline 
       20-25  & 25215 & 10081 & 8086  & 7048  & 14956 & 10259 & 0     \\
       25-30  & 44603 & 18655 & 16074 & 9874  & 10229 & 22281 & 12094 \\
       30-35  & 43981 & 14168 & 12446 & 17367 & 4993  & 16368 & 22620 \\
       35-40  & 19469 & 7236  & 5332  & 6901  & 2882  & 7757  & 8830  \\
       40-45  & 2380  & 1960  & 0     & 420   & 1554  & 826   & 0     \\
       45-50  & 445   & 445   & 0     & 0     & 445   & 0     & 0     \\
        \hline
    \end{tabular}
    \caption{In this table we bin the predictions from our fits in a synthetic population of 240,000 stars similarly to Table 3 in \citet{Winch24}. Predicted BH masses are separated into $5 M_\odot$ bins and total populations are listed in the second column. Columns 3-5 separate each population by its rotation rate, while Columns 6-8 are for groups of metallicity.}
    \label{T-IMF_Table_BH}
\end{table*}

\section{Discussion}
\label{s-Discussion}

We use this section to discuss some of the implications of our work, specifically in the context of mass loss enhancement by rotation and mechanical mass loss, or other forms of mass loss boost by massive stars. A lot of work has been done in the literature prior to this, though there is still much debate as to which methods are more appropriate for specific stars. Various assumptions between stellar evolution codes also contribute heavily to this discussion, often forming the foundation of key uncertainties and differences between model results. Exactly how we place our new results in the framework of the already existing literature is discussed below.

\subsection{High Rotation ($> 0.4 \Omega/\Omega_{\rm crit}$) and Low ($< 1/10$th $Z_{\odot}$) Metallicity}
\label{ss-Disc_HighRotLowZ}
Previous work using the Geneva code has provided valuable insight into how low metallicity, fast rotating massive stars might resolve their angular momentum excesses - specifically through mechanisms such as mechanical mass loss \citep{Krticka11,Georgy13,Murphy21}. Without using such mechanisms within MESA, alternatives have to be considered. As mentioned in Section \ref{s-Results}, the use of the rotation-induced mass loss enhancement of \citet{Yoon10} provides a mass loss boost which is more numerically stable than that of \citet{Maeder00B} in MESA, especially when using the mass loss recipe of \citet{Sab23}. 

Given the low mass-loss rates of low metallicity stars regardless of what the effective boost from either \citet{Maeder00B} or \citet{Yoon10} might be, the differences in final masses for these stars does not contribute as a major source of error on the predictions. However, this is mainly because the effect of these boosts is applied directly to the mass-loss rate itself. If the star were to experience supercritical mass loss due to reaching one of the respective limits (e.g. $\Omega$ or $\Omega \Gamma$ limits), then this would not necessarily be tied directly to the radiative mass-loss rate. 

\citet{Georgy13} addresses this by separating the radiative mass and angular momentum loss from the mechanical mass and angular momentum loss. It is likely that this is indeed the correct way to consider this form of evolution, however the exact method for doing so is still very much up for debate. Firstly, the results of \citet{Georgy13} are for stars with initial masses up to $15 M_{\odot}$, whereas the stars we examine in this work are at least three times larger. The additional proximity to the Eddington limit may mean that the $\Omega \Gamma$ limit is reached much sooner for these stars, which would imply supercritical mass loss would become a much more significant factor. 

One must also consider the differences between 1-D stellar evolution codes as a source of potentially significant error. As described in Section \ref{s-Methods}, the differences in the way MESA and GENEC handle rotational mixing and angular momentum transport are significant. The specific formula for how MESA handles angular momentum transport can be found in \citet{Paxton13}, using the diffusive mixing approximation of \citet{Endal78, Pinsonneault89, Heger00}, while GENEC follows both advection and diffusion described in \citet{Zahn92, Chaboyer92, Maeder98}. Although this paper does not aim to be a rigorous mathematical examination of the merits of either methodology, it is important to highlight several differences to provide meaningful comparisons to the literature. At the end of Section \ref{ss-Results-EffectsOfRotation} we briefly investigate this in the context of models in \citet{Ekstrom08} and \citet{Yoon12}. We found that, even for different implementations of rotational mixing, we can approximate the mass loss of models in \citet{Ekstrom12} by changing our own rotational mixing efficiencies. We identified that differences between the final masses of \citet{Yoon12} and \citet{Ekstrom12} were not caused solely by the use of the Spruit-Tayler dynamo or not, but instead a series of compound differences in the assumptions made such as the method of ``mechanical'' mass loss.

It is useful to compare our results to other mechanical mass loss models, especially at low metallicity where the effect of mechanical mass loss is the dominant form of mass loss over radiative winds. To accomplish this, we run a model where $M_{\rm ZAMS} = 120 M_\odot$ and $Z = 1/1000$th $Z_{\odot}$ and with an initial rotation rate of $0.4 \Omega_{\rm crit}$ to compare the differences between our implementation and that of the low metallicity grid in \citet{Sibony24}. The authors of \citet{Sibony24} calculate mechanical mass loss using the same method as \citet{Georgy13} which involves explicit calculation of the angular momentum of layers at the star to determine the angular momentum loss per unit time step.

\begin{figure}
    \centering
    \includegraphics[width=1.05\linewidth]{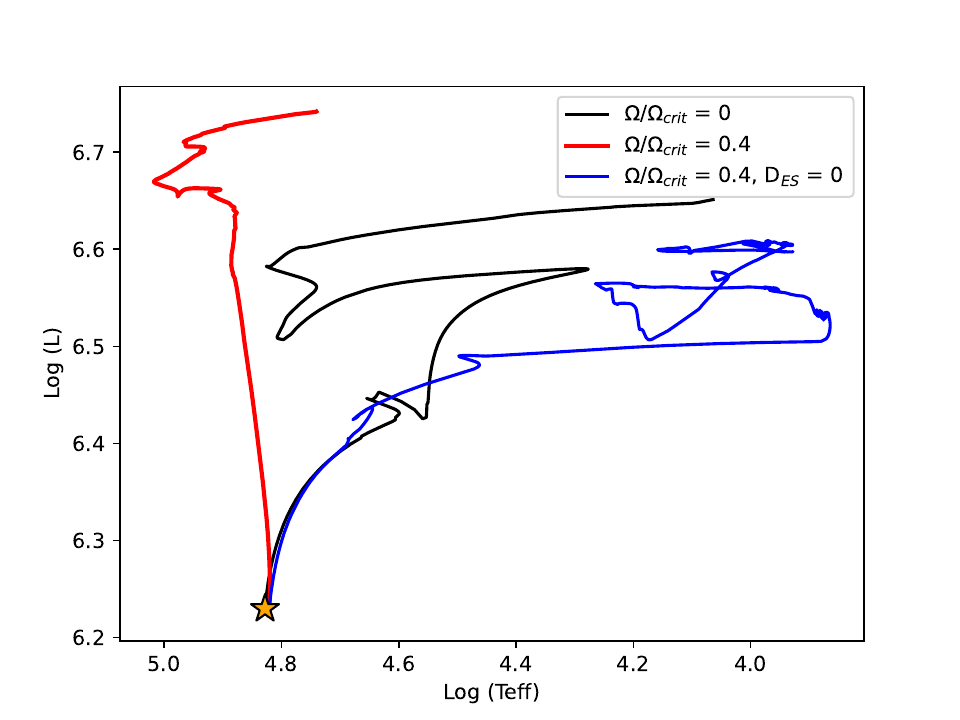}
    \caption{Hertzprung-Russel Diagram comparing several $120 M_{\odot}$ models. The red line tracks the evolution of a model with \OmOmC $= 0.4$, while the black line tracks a model with no rotation. The blue track is for a model with rotation, but with the diffusion coefficient for the Eddington-Sweet circulation \citep[$D_{\rm ES}$,][]{Heger05} set to 0. The orange star represents the ZAMS point for these models. All models are $Z = 1/1000$th $Z_\odot$ in order to compare with \citet{Sibony24}.}
    \label{Fig-SibonyComparison}
\end{figure}

Firstly, our model finishes core He burning with a final mass of $106.7 M_{\odot}$, which is $\sim 22 M_{\odot}$ greater than the value provided by \citet{Sibony24}, or approximately a $20 \%$ difference as a fraction of initial mass. Figure \ref{Fig-SibonyComparison} shows a set of three models we ran for comparison. For a fiducial model, we have the black line to demonstrate how MESA will evolve a 120 M star without rotation. If we add $0.4 \Omega_{\rm crit}$ rotation to the same model, MESA produces the red track. One key difference here is the trajectory of the model in the HRD - the MESA model with rotation turns hotter rather than cooler as it does in GENEC. In MESA, a similar behaviour can be replicated by setting $D_{\rm ES} = 0$, however it is critical to note that this does not entirely replicate a GENEC model as it does not strictly replicate the GENEC's advection implementation of angular momentum transport due to meridional circulation. 

One notable result from \citet{Sibony24} was the lack of any model reaching the WR stage due to mass loss alone. This is in direct contradiction to our models in Figure \ref{Fig-100M_HRDs_NoST_MechMdot}, where we see that, due to the high rotation rates of our models (\OmOmC $= 0.6$ as opposed to $0.4$ in their work), all of our models at $100 M$ enter the stripped star regime, though only the model at $1/10$th $Z_{\odot}$ experienced a typical Wolf-Rayet core He burning (CHeB) phase. Comparatively, all of the \citet{Sibony24} models ended their CHeB phase on the cooler side, at $\log(T_{\rm eff}) \approx 3.8$ for the high mass models. It is notable that GENEC models, when rotating, tend towards \textit{lower} temperatures \citep[e.g.][]{Meynet06} which is in stark contrast to MESA models which \textit{increase} in luminosity with more rotation, to the point where vertical, and hotter, evolution is expected from rotating models. 

The evolution of the fraction of critical velocity, \OmOmC, can be complicated, as the traditional view of ``spin-up'' is not the exclusive reason for the \OmOmC to increase. For example, the contraction of the star without any mechanism for which angular momentum is lost causes an increase in the \OmOmC. Alternatively, the critical velocity itself may decrease as per Equation \ref{Eq-MM00_vcrit_2} due to the Eddington parameter. As such, it is important to consider not only the surface velocity of a star, but also the critical velocity when considering the evolution of stars, and their mass-loss rates. As we show, at low metallicity, there can still be non-negligible amounts of mass loss for stars which have even moderate rotation, at these high initial masses. In fact, this was noticed in our very low metallicity models in \citet{Winch24}, and would imply that very low metallicity stars, which may have a more top heavy IMF \citep[][]{Bromm02,Vink18,Chruslinska20}, would be more likely to be subject to mechanical mass loss episodes. This in turn is likely to reduce the final black hole masses of the population - or at the very least, reduce the likelihood of reaching the potential maximum. 

\subsection{$M_{\rm crit}$}
\label{ss-Disc_MCrit}

Other authors before us have noted that there is a core mass limit for which one would observe dynamical instability from pair creation in their models such as \citet{Renzo17,Farmer19, Renzo20_LowerPISN, Marchant19}. However, this is normally examined in the context of initially stripped He stars and not those with a H-rich envelope (with the expectation that the envelope has been stripped from some other mechanism over the course of the star's evolution). We aim to define this as a critical core mass which is independent in changes to stellar evolution parameters. To this end, we investigate how these different physical mechanisms could change this limit. Within the limits of the parameters we test in \citet{Winch24}, we showed that both the He and Carbon-Oxygen cores did not change. However, this is no longer true for fast rotating stars which strip and thus neither is it true for stripped stars. Specifically with the critical He core mass, for fast rotator or stripped stars, the surface of the star becomes the Carbon-Oxygen core. As such, the He core mass is either 0 or is the mass of the CO core/final mass of the star depending on the definition of the He core mass.

\citet{Marchant20} demonstrated that rotation can have a significant impact on whether a star would be susceptible to PI or not. We test models without mass loss to isolate the effects of rotation without worrying about compound effects of rotation and to determine whether or not the criteria detailed by \citet{Marchant20} is a strong parameter in our determinations. For this we had several models of $38 M_{\odot}$ in order to be above the critical core mass ($36.3 M_\odot$) but not too far where the \citet{Marchant20} condition would not have a noticeable effect. For these models we disabled mass loss and rotation for the core H and core He burning phases of evolution, but set a large overshooting parameter to ensure the entire star was composed of a CO core. Once the model was He depleted and about to start core Carbon burning, we enabled rotation between $\Omega/\Omega_{\rm crit} = 0-0.6$. In this test, our rapid rotators were not unstable according to our criterion based on \citet{Stothers99}, which aligns with what is expected from \citet{Marchant20}. While we subscribe to the \citet{Marchant20} condition, our test shows that this is not a significant source of uncertainty for our $M_{\rm crit}$ experiment.

\section{Conclusions}
\label{s-Conclusions}

In this work we continue from \citet{Winch24} where we investigate the location of the PI boundary. Previously we were limited mostly to the low-rotation case without a method for dealing with stars which reach supercritical rotation rates. Now we push up our previous rotation limit by implementing mechanical mass loss and extending our $M_{\rm crit}$ experiment up to $\Omega = 0.8 \Omega_{\rm crit}$, confirming that the PI limit remains a constant critical CO core mass, though we see that the He core mass is reduced as the star becomes stripped. Finally, we run a grid of models across different rotation rates to map the potential BH progenitors in this space. We find that the amount of mass lost by mechanical methods is highly dependent on the ability of a model to remove excess angular momentum, nominally through stellar winds. This itself has a strong metallicity dependence, which we show leads to the conclusion that low metallicity stars with rotation are likely to have at least one period of mechanical mass loss during their evolution, typically at a contraction phase such as the Henyey Hook, which occurs at the end of the main sequence.\\

To summarise our results:\\

\begin{itemize}
    \item  At high metallicity models lose a significant amount of mass through stellar winds, and rotationally enhanced stellar winds \citep[][]{Maeder00B}. There is no mechanical mass loss as the wind mass loss is sufficient to strip angular momentum. The ST dynamo has the most significance here on the final mass of models.\\
    \item Low metallicity stars lose less mass in total than high metallicity, but a large fraction of the mass loss is due to mechanical mass loss - this fraction increases the lower in metallicity the star is. These models are also more likely to undergo CHE, and thus reduce the final BH mass as $M_{\rm final} = M_{\rm core}$, so the limit on the final mass for a BH is the critical core mass. \\
    \item The ST dynamo has almost no significance on the final mass at low $Z$ ($Z < 1/10$th $Z_\odot$). This disagrees with \citet{Yoon12} who states that the difference between their models and those of \citet{Ekstrom08} is exclusively due to the presence or lack of an ST dynamo. \\
    \item When accounting for only the high rotation models, the critical core mass for is $M_{\rm CO, crit, high \Omega / \Omega_{\rm crit}} \approx 35 M_{\odot}$ which is within the uncertainty range of our critical core mass in \citet{Winch24}. The initial mass of models at the boundary for high rotation increases as these models need the extra mass to lose compared to lower rotation models. The maximum final mass of these models is exactly the critical core mass as they are fully mixed (so the core is the entire star). This is within the error bars of \cite{Winch24} and is lower due to the CO critical core masses at 0.4, 0.5 being notably lower as this is where the models transition from H rich to stripped. \\
    \item $M_{\rm He}$ is no longer valid for these models due to the stars becoming fully mixed CO cores - thus $M_{\rm He}$ is seen to “decrease”. \\
    \item We have also created two fits for $M_{\rm CO}$ and $M_{\rm final}$ (and one for $M_{\rm He}$ for completeness with \cite{Winch24}), and subsequently created a population of BH progenitors using the same mechanism of \cite{Winch24}. We notice much reduced masses in comparison to the low rotation case, with a peak starting at $\sim 27 M_\odot$. The steep drop observed after $36.3 M_\odot$ is due $M_{\rm crit}$ maximum, where models are mostly chemically homogenous. Above this, there are a few stars which still have a small He shell above the CO core. \\
    \item The steep drop observed after $ \sim 36.3 M_\odot$ coincides with the bump feature in the BH mass distribution of GW events as observed by LIGO/VIRGO \citep[][]{Abbott21c}.
\end{itemize}

\section*{Acknowledgements}

The authors acknowledge MESA authors and developers for their continued revisions and public accessibility of the code, and the referee for constructive comments. EW is funded by ST/W507925/1. JSV and ERH are supported by STFC funding under grant number ST/V000233/1 (PI Vink). The authors would like to thank the referee for their helpful comments on the manuscript.

\section*{Data Availability}

Input files for variables will be made publicly accessible via the MESA marketplace.


\bibliographystyle{mnras}
\bibliography{references} 



\appendix
\onecolumn
\section{Table of Models}
\label{Ap-TableOfModels}

\begin{tabular}{ p{1cm}p{1cm}p{1cm}p{1cm}p{1cm}|p{1cm} p{1cm}|p{1cm}p{1cm}p{1cm}}
\hline
\hline
 $Z / Z_{\odot}$ & $M_\text{i}$ & \aOv & \OmOmC & $D_{\rm ST}$ & $M_{\text{TAMS}}$ & $M_{\text{He}}$ & $M_{\text{f}}$ & $M_{\text{CO}}$ & $M_{\text{He}}$ \\  
\hline
0.1 & 50 & 0.1 & 0.4 & 0 & 44 & 23 & 31 & 0 & 27\\
0.1 & 50 & 0.1 & 0.6 & 0 & 42 & 28 & 27 & 23 & 27\\
0.1 & 50 & 0.1 & 0.8 & 0 & 41 & 30 & 25 & 22 & 25\\
0.01 & 50 & 0.1 & 0.4 & 0 & 47 & 33 & 41 & 0 & 39\\
0.01 & 50 & 0.1 & 0.6 & 0 & 46 & 42 & 41 & 36 & 41\\
0.01 & 50 & 0.1 & 0.8 & 0 & 46 & 42 & 41 & 35 & 41\\
0.001 & 50 & 0.1 & 0.4 & 0 & 49 & 42 & 45 & 39 & 45\\
0.001 & 50 & 0.1 & 0.6 & 0 & 48 & 44 & 48 & 0 & 44\\
0.001 & 50 & 0.1 & 0.8 & 0 & 48 & 44 & 42 & 36 & 42\\
\hline
0.1 & 60 & 0.1 & 0.4 & 0 & 53 & 28 & 36 & 0 & 32\\
0.1 & 60 & 0.1 & 0.6 & 0 & 50 & 34 & 32 & 28 & 0\\
0.1 & 60 & 0.1 & 0.8 & 0 & 49 & 35 & 28 & 24 & 28\\
0.01 & 60 & 0.1 & 0.4 & 0 & 56 & 41 & 54 & 0 & 48\\
0.01 & 60 & 0.1 & 0.6 & 0 & 55 & 51 & 49 & 43 & 49\\
0.01 & 60 & 0.1 & 0.8 & 0 & 55 & 51 & 48 & 42 & 48\\
0.001 & 60 & 0.1 & 0.4 & 0 & 58 & 51 & 53 & 47 & 53\\
0.001 & 60 & 0.1 & 0.6 & 0 & 58 & 53 & 51 & 44 & 51\\
0.001 & 60 & 0.1 & 0.8 & 0 & 57 & 53 & 50 & 43 & 50\\
\hline
0.1 & 70 & 0.1 & 0.4 & 0 & 61 & 34 & 43 & 0 & 37\\
0.1 & 70 & 0.1 & 0.6 & 0 & 58 & 39 & 37 & 32 & 0\\
0.1 & 70 & 0.1 & 0.8 & 0 & 56 & 41 & 30 & 26 & 0\\
0.01 & 70 & 0.1 & 0.4 & 0 & 66 & 45 & 63 & 0 & 53\\
0.01 & 70 & 0.1 & 0.6 & 0 & 64 & 59 & 57 & 50 & 57\\
0.01 & 70 & 0.1 & 0.8 & 0 & 64 & 59 & 56 & 50 & 56\\
0.001 & 70 & 0.1 & 0.4 & 0 & 68 & 60 & 62 & 0 & 62\\
0.001 & 70 & 0.1 & 0.6 & 0 & 67 & 63 & 59 & 52 & 59\\
0.001 & 70 & 0.1 & 0.8 & 0 & 66 & 62 & 58 & 51 & 58\\
\hline
0.1 & 80 & 0.1 & 0.4 & 0 & 69 & 39 & 51 & 0 & 43\\
0.1 & 80 & 0.1 & 0.6 & 0 & 65 & 45 & 44 & 0 & 0\\
0.1 & 80 & 0.1 & 0.8 & 0 & 63 & 47 & 33 & 29 & 0\\
0.01 & 80 & 0.1 & 0.4 & 0 & 75 & 52 & 72 & 0 & 60\\
0.01 & 80 & 0.1 & 0.6 & 0 & 73 & 68 & 65 & 0 & 65\\
0.01 & 80 & 0.1 & 0.8 & 0 & 72 & 67 & 63 & 0 & 63\\
0.001 & 80 & 0.1 & 0.4 & 0 & 78 & 69 & 70 & 0 & 70\\
0.001 & 80 & 0.1 & 0.6 & 0 & 76 & 72 & 67 & 0 & 67\\
0.001 & 80 & 0.1 & 0.8 & 0 & 75 & 71 & 66 & 0 & 66\\
\hline
0.1 & 90 & 0.1 & 0.4 & 0 & 77 & 45 & 58 & 0 & 48\\
0.1 & 90 & 0.1 & 0.6 & 0 & 73 & 50 & 53 & 0 & 53\\
0.1 & 90 & 0.1 & 0.8 & 0 & 70 & 52 & 35 & 31 & 0\\
0.01 & 90 & 0.1 & 0.4 & 0 & 84 & 57 & 81 & 0 & 64\\
0.01 & 90 & 0.1 & 0.6 & 0 & 82 & 76 & 72 & 0 & 72\\
0.01 & 90 & 0.1 & 0.8 & 0 & 81 & 76 & 70 & 0 & 70\\
0.001 & 90 & 0.1 & 0.4 & 0 & 88 & 73 & 80 & 0 & 79\\
0.001 & 90 & 0.1 & 0.6 & 0 & 86 & 81 & 75 & 0 & 75\\
0.001 & 90 & 0.1 & 0.8 & 0 & 84 & 79 & 74 & 0 & 74\\
\hline
0.1 & 100 & 0.1 & 0.4 & 0 & 85 & 50 & 63 & 0 & 54\\
0.1 & 100 & 0.1 & 0.6 & 0 & 80 & 56 & 61 & 0 & 61\\
0.1 & 100 & 0.1 & 0.8 & 0 & 77 & 58 & 37 & 33 & 0\\
0.01 & 100 & 0.1 & 0.4 & 0 & 93 & 65 & 90 & 0 & 73\\
0.01 & 100 & 0.1 & 0.6 & 0 & 91 & 84 & 79 & 0 & 79\\
0.01 & 100 & 0.1 & 0.8 & 0 & 90 & 84 & 77 & 0 & 77\\
0.001 & 100 & 0.1 & 0.4 & 0 & 97 & 75 & 97 & 0 & 75\\
0.001 & 100 & 0.1 & 0.6 & 0 & 95 & 90 & 83 & 0 & 83\\
0.001 & 100 & 0.1 & 0.8 & 0 & 93 & 88 & 82 & 0 & 82\\
\end{tabular}


\bsp	
\label{lastpage}
\end{document}